\documentclass[12pt,preprint]{aastex}

\shorttitle{M31 and M33 Luminous Stars}
\shortauthors{Humphreys et al. }

\begin{document}

\title{Luminous and Variable Stars in M31 and M33. IV. Luminous Blue Variables, Candidate LBVs, and the B[e] Supergiants; How to Tell Them  Apart\altaffilmark{1}}

\author{
Roberta M. Humphreys\altaffilmark{2}, 
Michael S. Gordon\altaffilmark{2},
John C. Martin\altaffilmark{3},
Kerstin Weis\altaffilmark{4},
and 
David Hahn\altaffilmark{2},
}

\altaffiltext{1}  
{Based  on observations  with the Multiple Mirror Telescope, a joint facility of the Smithsonian Institution and the University of Arizona and on observations obtained with the Large Binocular Telescope (LBT), an international collaboration among institutions in the United
States, Italy and Germany. LBT Corporation partners are: The University of
Arizona on behalf of the Arizona university system; Istituto Nazionale di
Astrofisica, Italy; LBT Beteiligungsgesellschaft, Germany, representing the
Max-Planck Society, the Astrophysical Institute Potsdam, and Heidelberg
University; The Ohio State University, and The Research Corporation, on
behalf of The University of Notre Dame, University of Minnesota and
University of Virginia.
} 

\altaffiltext{2}
{Minnesota Institute for Astrophysics, 116 Church St SE, University of Minnesota
, Minneapolis, MN 55455; roberta@umn.edu} 

\altaffiltext{3}
{University of Illinois Springfield, Springfield, IL 62703}

\altaffiltext{4}
{Astronomical Institute, Ruhr-Universitaet Bochum, Germany,}

\begin{abstract}
In this series of papers  we have presented the results of a spectroscopic survey of 
luminous and variable stars in the nearby spirals M31 and M33. In this paper, we present spectroscopy 
of 132 additional luminous stars, variables, and emission line objects. Most of the stars have  
emission line spectra, including LBVs and candidate LBVs, Fe II emission line stars and the B[e] 
supergiants, and the warm hypergiants. Many of these objects are spectroscopically similar and are 
often confused with each other.  With this large spectroscopic data set including various types of 
emission line stars, we examine their similarities and differences and propose the following criteria
that can be used to help distinguish these stars in future work:
1.  The B[e] supergiants have emission lines of [O I] and [Fe II] in their spectra. Most of the 
spectroscopically confirmed sgB[e] stars also have warm circumstellar dust in their SEDs. 
2. Confirmed LBVs do not have the [O I] emission lines in their spectra. Some LBVs have 
  [Fe II] emission lines, but not all. Their SEDS shows free-free emission in the near-infrared but 
  {\it no evidence for warm dust}. Their most important and defining characteristic is 
  the S Dor-type variability.  3.  The warm hypergiants spectroscopically resemble both the LBVs in their eruption or dense wind state 
  and the B[e] supergiants. However, they are very dusty. Some have [Fe II] and [O I] emission in 
  their spectra like the sgB[e] stars, but can be distinguished by 
  their absorption line spectra characteristic of A and F-type supergiants. In contrast, the 
  B[e] supergiant spectra  have strong continua and few if any apparent absorption lines.\\ 
Candidate LBVs should share the spectral characteristics of the confirmed LBVs with low outflow 
velocities and the lack of warm circumstellar dust. Based on these guidelines, we suggest that  the Fe II 
emission line stars (no [O I], low outflow velocities, and no dust), should be considered LBV 
candidates.
\end{abstract} 


\section{Introduction -- The Diverse Population of Luminous Emission Line Stars}

The upper HR diagram is populated by a diversity of luminous and variable 
stars of different types  many of which are distinguished by 
their emission line spectra, and evidence for stellar winds and mass loss. 
The different groups or types include the Luminous Blue Variables (LBVs), B[e]
supergiants, the Wolf-Rayet stars of different types, and more generic hot emission line stars. Many of these stars 
occupy the same regions of the HR diagram, and they may or may not be related. 
They could be stars of similar initial mass but in different stages of their evolution or have experienced different mass loss histories. This diversity is one
of the challenges to understanding massive stars, their evolution, and eventual fate.

In this series of papers  we have presented the results of a spectroscopic survey of 
luminous and variable stars in the nearby spirals M31 and M33. In Paper I \citep{RMH13}
we discussed a small group of very luminous intermediate temperature supergiants, the warm hypergiants, 
and showed that they were likely post-red supergiants. In Paper II \citep{RMH2014b}, we reviewed  the spectral characteristics,  spectral energy distributions (SEDS), circumstellar
ejecta, and  mass loss for 82 luminous and variable stars including the 
confirmed LBVs, candidate LBVs, and other emission line stars. 
We found that many of these stars have warm circumstellar dust including several of 
the stars with Fe II and [Fe II] emission lines, but concluded that the confirmed LBVs in M31 and 
M33 do not.  Interestingly, the confirmed LBVs have 
relatively low wind speeds even in their hot, quiescent or visual minimum state 
compared to the 
B-type supergiants and Of/WN stars which they spectroscopically resemble. 

The final state of the most massive stars as core collapse SNe is now in question. Smartt et al. (2009, 2015) have suggested an upper mass 
limit of $\approx$ 18 M$\odot$ for the red supergiant progenitors of 
the Type II SNe, while \citet{Jennings} find
a lack of massive progenitors in M31 and M33 and suggest an upper mass 
of 35-45 M$\odot$
In our Paper III, \citet{Gordon} presented  a  comprehensive spectroscopic survey of the yellow supergiants. Based on spectroscopic 
evidence for mass loss and the presence of circumstellar dust in their SEDs, 
we concluded that $30-40\%$ of the yellow supergiants are likely in a post-red 
supergiant state. Comparison with evolutionary tracks shows that these 
mass-losing, post-RSGs have initial masses between $20-40\,M_{\odot}$ suggesting
that red supergiants in this mass range evolve back to warmer temperatures before their terminal state. This is a significant result given the evidence that the most massive stars may not end their lives as supernovae and the recent announcement of a possible failed SN  from  a  red supergiant candidate \citep{Adams}.  

In this paper (Paper IV), we present spectroscopy of 132 additional luminous stars, variables, 
and emission line objects. Most of the stars have  
emission line spectra, including LBVs and candidate LBVs, Fe II emission line stars and the B[e] supergiants, and the warm hypergiants. Many of these objects are spectroscopically similar and are often confused with each other. For example, stars with Fe II and [Fe II] emission are sometimes assumed to be LBV candidates.  With this large spectroscopic data set including various types of emission line stars, we
can examine their similarities and differences. We briefly describe our 
new observations in the next section. In the following sections, we discuss  
their spectroscopic and
photometric properties, their SEDs, and presence or lack of 
circumstellar dust.  In the last section we summarize their characteristics  
and suggest guidelines that can be used to help distinguish or separate these 
stars in future work.

\section{New Observations} 

Our target selection and observations with the MMT/Hectospec \citep{Fab}  and LBT/MODS1
were  described in Paper I. Our subsequent new spectra 
are listed here in Table 1. 
In Paper II, we assigned our targets to six different groups according to their 
spectroscopic and photometric characteristics described there, and we follow
that designation in  this paper.  All of the hot stars for which we have 
new spectra are listed in  Table 2 in order of
Right Ascension with their  position\footnote{We use the impoved positions from the revised catalogs for M31 and M33 from \citet{Massey16}}, group type, spectral type where appropriate, and alternate names or designations. The yellow supergiants (YSGs) are listed in Paper III. The M33C designation in Table 2 comes from an unpublished H$\alpha$ survey by Kerstin Weis. The  ``V-'' notation is used for stars selected 
from the emission line survey of M33 by \citet{Valeev}.
In this paper,  we use the galaxy name and the RA of the star as its designator for brevity and to save space in the later tables. A shortened version is shown here. The full length table is available in the on-line edition. 

We repeated spectra of several  stars discussed in Papers I and  II 
to look for variability including the 
confirmed and candidate LBVs,  the hypergiants,  plus several 
emission-line stars. These
stars are listed separately in Table 3 following the same format as Table 2. 
Any noteworthy spectroscopic changes are noted in the following  sections. 
We also continued to photometrically monitor many of these stars at the Barber Observatory as described in \citet{RMH15}. 
This multi-color photometry is available for most of the stars discussed in this series of papers at 
 http://etacar.umn.edu/LuminousStars/.

The
visual photometry from \citet{Massey06} was cross-identified with the  near- and mid-infrared photometry 
from  2MASS \citep{Cutri},  the {\it Spitzer} surveys of M31 \citep{Mould} and M33 \citep{McQ,Thom}, and WISE \citep{Wright}. The first 10 entries are shown in Table 4 and the full table is available in 
the on-line edition. Table 4 also includes information on variability for the M31 stars from the DIRECT survey (see \citet{Kaluzny}, and for M33 stars  from \citet{Hartman} and from \citet{BurgPhD}. The Burggraf compilation includes data from  several sources from $\sim$ 1920 to the present, but  most are since 1970. We have also include information on variability from our own monitoring of individual stars since 2010.  
 The blue and red spectra in FITS format of all of the stars observed with the MMT/Hectospec are available at etacar.umn.edu/LuminousStars. The year of observation
 is also included in Tables 2 and 3 to aid in locating the spectra in the on-line database.

\section{ The Luminous Blue Variables and Candidate LBVs}

The LBV/S Dor variables exhibit extended periods of enhanced mass loss distinguished by a specific spectroscopic and photometric variability \citep{HD94,RMH16}. During their maximum  (visual) light or dense wind stage their spectra transition from a hot star
with mass  loss to resemble a cooler A or F-type supergiant spectrum produced by the optically thick wind. Given the infrequency of these high mass loss episodes, i.e.  the S Dor variability, few confirmed LBVs are known in M31 and M33.  

The confirmed LBVs including the recently identified J004526.62+415006.3 \citep{Shol,RMH15}  are listed
in Table 5  with the few candidates discussed here. Their spectra from  our observations in 2010 were described  in Paper II. These LBVs and candidates 
have been observed again in 2013 -- 2015 to look for any significant 
spectroscopic changes. In this section we briefly discuss those LBVs that have 
shown some  spectroscopic variability since 2010. The short term  variability 
noted here in the spectra for several of the LBVs is very likely typical of 
their instability and the resulting variations in their winds and mass loss,  
even when they are in  their quiescent state.  

The visual spectrum of {\it AE And} has always shown strong emission lines
of Fe II and [Fe II] \citep{RMH75}. In Paper II we noted that the strong 
H and He I emission lines had weakened compared to an earlier spectrum from 2004. Below 4100{\AA} these lines  were in absorption, and their relative strengths
suggested a corresponding B2 - B3 spectral type for the underlying star. Thus the  wind appeared to be weakening, but its spectrum from 2013 shows that although the absorption lines were still present, the P Cygni emission had re-appeared, and in the 2014 spectrum, the absorption has been completely replaced by emission. These variations in the spectrum  indicate continued instability and variation in the mass loss even though AE And is in its quiescent state.  

{Var A-1} in M31 has  prominent Fe II and [Fe II] emission with H and He I P Cygni profiles and N II absorption lines. In the past (Paper II), it has shown variation in the strength of the P Cygni profiles, but its spectra from 2010, 2013, and 2015 
show no substantial change. 

{\it AF And} and {\it Var 15} in M31, in quiescence, have spectra like the Of/late WN stars. Both showed substantial change in the few years from  2010 to 2013 or 2015. The P Cyg absorption minima in the Balmer lines in AF And are 
much stronger in 2013 and 2015, although there is 
no similar change in the He I P Cyg profiles. 

In Var 15, there was no spectral variability between 2010 and 2013, but between 2013 and 2015,
the {\it He I and [Fe II] emission lines disappeared and the N II emission line greatly weakened} as shown in Figure 1. Our photometric 
monitoring at the Barber Observatory shows that Var 15 brightened by 0.4 -- 0.5 mags in the BV and R  bands from
Sept 2013 to Sept 2014 and continued to brighten by another 0.1 mag from 2014 to 2015. During the same time its $B-V$ color however reddened only slightly from  0.4 to 0.5 mag.  Thus the spectroscopic and photometric changes are correlated. The disappearance of the He I emission suggests a decrease in the apparent temperature supported by the somewhat redder color. Previous photometry
\citep{Szeif} from 1992 showed Var 15 significantly brighter but with a similar color while photometry in \citet{Massey06} from circa 2000 shows a fainter star with much bluer $B-V$ 
color   of -0.01 mag.   Clearly, Var 15 has undergone some significant changes in its wind that have largely gone unnoticed. It should be monitored more closely. 

The new LBV in M31, {\it J004526.62+415006.3}, currently in ``eruption'', developed enhanced
emission in the Fe II lines in the 5000 - 5600{\AA} region in the 2015 spectrum relative to 2013, although the absorption line spectrum  is unchanged. Recent photometry shows that it has faded about 0.25 mag but the color is unchanged.

{\it Var 83} in M33 has a spectrum very much like Var A-1 with N II absorption lines, Fe II and [Fe II] emission lines, and  with strong P Cygni profiles in the Balmer lines.  Stronger P Cygni emission profiles in the He I lines appeared between 2010 and 
2013, but there was no significant change in the 2014 spectrum. However, subsequently Var 83 brightened 
0.5 mag in V between November 2014 and October 2015, but its color did not change. It has remained at $\cong$ 15.8 mag since. 

{\it Var B and Var 2} in M33  have spectra like the Of/late WN type stars with nitrogen emission and He I P Cygni profiles. Var 2 also has He II the  $\lambda$4686 in emission ( Figure 5 in Paper II). No changes are noted in their spectra from 
2010 to 2013 and 2014, but the Fe II and [Fe II] emission lines in Var B  disappeared between  2004 to 2010 (Figure 2 in Paper II). 

{\it Var C} in M33 is currently  in ``eruption'' \citep{RMH2014a,Burg15}. No 
additional spectra were obtained after 2013, but the available photometry shows that it is still in its dense wind state although,  it has faded about 0.5 mag between Nov. 2015 and Sept 2016, and the $B-V$ color has shifted from 0.2 to 0.0. So, it is probably transitioning back to its hotter, quiescent state..     

This brief survey of the confirmed LBVs in their quiescent state revealed spectroscopic variations which in most cases are small and  due to variability  in their 
winds and mass loss. This is not surprising for stars known for instability. 

In Paper II we showed that the SEDs of the  confirmed LBV/S Dor variables do not have any evidence for warm dust  in the  near- and mid-infrared. 
We also 
found that the LBVs have relatively slow winds, even in their quiescent state, 
typically $\approx$ 100 -- 200 km s$^{-1}$, compared to the winds of normal hot supergiants and the Of/WN stars.  These characteristics can also be used to identify LBV candidates and separate them from other 
emission line stars.  

\subsection{LBV Candidates} 

Numerous stars have been suggested as candidate LBVs in the literature \citep{Massey07,Clark12,Massey16,Shol}. In Paper II, we also recognized several stars as potential LBVs.  Our spectra of several of the stars that have been suggested as LBV 
candidates are described here.  

B526 \citep{HS80} (= M33C-7292) has been considered a candidate LBV based on its spectral appearance \citep{Clark12}, but it was also suspected to be more than one star \citep{Mont}. HST/WFPC2 images\footnote{HST/WFPC2 images of the B526 field are available from three epochs and cover the wavelength region from the near UV (F170W)  to red (F814W). UBVRI magnitudes were measured  for the two stars using DOLPHOT.} (Figure 2) show that B526 is two stars of comparable brightness about  one arcsec apart. For this reason we obtained long slit spectra with the LBT/MODS1 in good seeing which easily separated the two stars. The NE star has a spectrum of a hot supergiant with Fe II emission, weak [Fe II] emission, H emission with P Cygni profiles, and He I absorption lines thus resembling many LBVs in their
quiescent state. The SW component has the normal absorption line spectrum of 
a  B5-type supergiant.  The blue and red spectra of these two stars are shown
 in Figure 3 and photometry for the two stars measured from the HST images is
 included in Table 4.  

The infrared photometry from 2MASS and Spitzer/IRAC is for the combined light of the two stars. Although they cannot be separated, their long wavelength SED 
clearly shows a lack of warm dust in the system similar to LBVs. However its wind velocity of 370 km s$^{-1}$, measured from the P Cyg minima in the Balmer lines is significantly higher than in the confirmed LBVs. 
 As noted later, its red spectrum has the [O I] emission at $\lambda$6300, a characteristic of the B[e] supergiants,  not present in the spectra of LBVs.
Thus as we will discuss later,  B526NE may be a normal, mass losing hot supergiant or perhaps a transition star 
possibly related to the B[e] supergiants, \S {4}.

In Paper II, we suggested that the spectrum of M31-004425.18 had changed from an early A-type spectrum described by \citet{Massey07} who called it a cool LBV candidate,
to a much hotter star, based on it absorption line spectrum described in Paper II. M31-004425.18 should definitely be
listed as an LBV candidate if not a true LBV based on its apparent spectroscopic
change like the S Dor variables, although it has a relative low luminosity compared to the known LBVs and lies below the S Dor 
instability strip; see Figure 12 in Paper II. Our spectra from  2013 and 2015 compared to the 2010 spectrum show  small variations 
in the strength of the He I emission indicative of variability in the wind and mass loss typical of LBVs in quiescence. 

UIT008 (J013245.41+303858.3) is an Of/late WN star. In Paper II, we classified it O7-O8f I and 
\citet{Massey98} called it Ofpe/WN9\footnote{Additional spectra, observed by Massey and collaborators, 
from 2010 and 2012  are available at the CfA Optical/Infrared Archive. They do not show any spectral variability.However, the P Cygni profiles and asymmetric wings are not visible in these lower resolution spectra.} H$\alpha$ and H$\beta$ have asymmetric profiles with a broad wing to the red suggestive of electron scattering in the wind.  
UIT008 also has a very low outflow velocity for a star of this type of only -159 and -138 km s$^{-1}$ measured respectively from the H and He I emission lines (Paper II). It is on this basis that we call it an 
LBV candidate. See \S {7} this paper for a discussion of the Of/late WN stars as candidate LBVs.

M31-004051.59 has been suggested to be an LBV or candidate LBV by \citet{Shol} and \citet{Massey07,Massey16}. We discussed its spectrum and SED in Paper III and concluded that its nature is somewhat ambiguous. It is spectroscopically similar to an LBV at maximum light in the dense wind state, and to the mass losing post RSGs discussed in Paper III. It has the absorption line spectrum of an early A-type supergiant with strong P Cygni profiles and broad wings at H$\alpha$  and 
H$\beta$. The prominent P Cygni profiles in the  multiplet 42  Fe II lines  
in the 2013 spectrum  weakened in 2015 and Fe II emission lines appeared at $\lambda$$\lambda$5234,5276,5284.  Its SED shows a small excess in the near-infrared due to 
free-free emission, but  no evidence for circumstellar dust like LBVs.

\citet{Massey16} listed  several stars with 
Fe II and [Fe II] emission line spectra as LBV candidates. Fe II emission 
lines are common in the confirmed LBVs, but as we discuss in the next 
section many of the   stars listed in their Tables 3 and 4 are actually B[e] supergiants.
The B[e] supergiants   also have significant circumstellar dust not observed in the confirmed LBVs.

\section{Fe II Emission-line Stars and the Supergiant  B[e] Stars}

We emphasized in Paper II, that the presence of Fe II emission lines is ubiquitous in astronomical spectra. Fe II emission is observed in many different types of stars with a wide range of luminosities and temperatures and in different evolutionary states. We defined the Fe II emission-line group  as those stars with 
a strong blue continuum,  
with strong hydrogen emission, Fe II emission,  and a lack or dearth of  absorption lines such that the nature of the underlying star was not clear. 

A significant subset of the Fe II emission line group known as  B[e] stars, also show [Fe II] in emission 
plus other forbidden lines such as [O I] and [Ca II].  In a recent paper on a spectroscopic survey of emission line stars, \citet{Aret2016} 
designated [O I] 
$\lambda$$\lambda$6300,6364 emission as one of the characteristics of the B[e] class. 
They also noted that the  [Ca II] doublet at $\lambda$$\lambda$7291,7324, an indicator of circumstellar gas, while  not 
present in all of  the B[e] stars, is more common in those of high mass. 
The upper level of the transition that produces the  [Ca II] lines is the lower energy state of the near-infrared Ca II 
triplet.  Thus [Ca II] emission implies emission in the Ca II triplet which we  confirmed for the 
stars with far red spectra\footnote{The Ca II triplet in emission is  present in the far red spectra of M33C-15731 and M31-004417.1 that were observed with MODS on the LBT (Paper II)}.  The ratios  of the Ca II and [Ca II] lines can then be used to derive the
electron densities in the circumstellar ejecta, see Paper I and II.  Furthermore the Ca II triplet in emission is 
produced by over-population of the upper level for the Ca II H and K lines in absorption. This in turn suggests that the 
underlying star must be relatively cool; perhaps of spectral type late B to A or even later.

The supergiant B[e] stars 
(hereafter sgB[e]) are of special interest in this study not only because of 
their luminosities, but also  with their  Fe II and [Fe II] emission lines, they are often mistaken for LBVs or LBV candidates. 
In Table 5 we list all of the stars in the Fe II emission line group from this
paper and from  Paper II plus two in M31 from the list of candidates in \citet{Massey16} with a summary of the permitted and forbidden emission lines observed in their spectra; Fe II, [Fe II], He I, [O I], [Ca II], the Ca II triplet , and the O I triplet at $\lambda$7774{A}. All of these stars have hydrogen 
emission.  We classify the stars with both [Fe II] and [O I] emission as 
B[e] stars.  All of the B[e] stars in M31 and M33 are luminous stars, i.e supergiants,  hence we adopt the sgB[e] 
designation and use it in Tables 2 and 3.  
Based on this criterion, eighteen of the 28 stars  in Table 5 are B[e] supergiants. Most also have significant excess infrared radiation due to warm dust, over and 
above what would be expected from free-free emission from their stellar winds,  \S {6}. 
The remaining ten stars are  Fe II emission line stars and have neither 
[Fe II] nor [O I] emission lines in their spectra. They also lack circumstellar dust like the LBVs.
Fifteen of the stars included as LBV candidates by \citet{Massey16} are B[e] supergiants.  
Repeat spectra were obtained for five of the sgB[e] stars and one Fe II emission line star, and  no changes were noted.

For comparison, we include the LBVs in Table 5. The notation for the emission lines refers to the LBV 
when in its quiescent state
for which we use our spectra from 2010, unless the same lines showed a significant change in previous or later  spectra such as for 
M31 Var 15 in 2015 when the [Fe II] were no longer present and Var B in 2004 (Paper II). This emission line summary shows a distinct difference with
respect to the B[e] supergiants. Most  of the LBVs have  [Fe II] and/or Fe II emission lines but none have [O I] in emission
or [Ca II].  The same spectrosopic difference is shown by the candidate LBVs M31-004425.18 and M31-004051.59  discussed in \S {3.1}. The spectrum of  B526NE, however,  does have the [O I] line like the sgB[e] stars. Its immediate neighbor B526SW  does not, so we can rule out contamination by emission nebulosity or a residual night sky line as a possible explanation.

\section{The Warm Hypergiants}

In Paper I we identified seven stars that we labeled warm hypergiants based on their high luminosity A and F-type absorption line spectra, the presence of strong Balmer emission with broad wings and P Cygni profiles, and the Ca II triplet and [Ca II] lines in emission. The ratios of the Ca II and [Ca II] lines  
provide information on the electron densities in the cicumstellar gas which range from  $\sim$ 1 to 4 $\times$ 10$^{7}$ cm$^{-3}$. They all have significant 
near- and mid-infrared excess radiation due to free-free emission and/or thermal emission from dust. Six of the stars have very dusty environments and on that basis, we suggested that they are post-RSG candidates, see also Paper III.
B324, the visually most luminous star in M33, does not have CS dust, and therefore may be an example of a massive star near the upper luminosity boundary in the HR diagram, evolving towards cooler temperatures.
In Paper III, we recognized another warm hypergiant, M31-004621.05, that had previously been considered an LBV candidate
\citep{Massey07}. Its spectrum and SED are in Paper III and its spectrum is also shown here in Figure 4. 

Repeat spectroscopy was obtained for all of the warm hypergiants. None of them  showed any variability except for M31-004322.50. 
There was no change from 2010 to 2013, but the prominent Fe II P Cygni emission profiles in multiplet 42 were much weaker in 2015 and the Fe II emission lines with P Cygni profiles in the red spectrum  were replaced by absorption lines signifying possible changes in its wind and a decrease in the density of its circumstellar gas.   

The warm hypergiants are listed in Table 6 with their emission line summary for comparison with the B[e] supergiants and the LBVs in
Table 5. In many respects their emission line characteristics plus the presence of circumstellar dust overlap with the properties 
of the sgB[e] stars. 
A couple of hypergiants have both [Fe II] and  [O I] emission. We caution however about the presence of weak [O I] which could be from surrounding nebulosity or a residual from night sky subtraction especially if the second, weaker line at $\lambda$6364 is not clearly present.  But the presence of absorption 
lines in the warm hypergiants typical of intermediate temperature supergiants, clearly separates them from the sgB[e] stars whose spectra lack  apparent 
absorption lines. Furthermore, the lack of He I emission together with  the O I $\lambda$7774 triplet in absorption in the hypergiants  
also indicates a significant temperature difference between them  and the sgB[e] stars. The [Ca II] doublet in 
emission and by implication the Ca II triplet in emission is another characteristic that some of the B[e] supergiants 
share with the hypergiants. \citet{Aret12} noted this spectroscopic similarity in several sgB[e] stars in the LMC and 
suggested that they
may be candidates for blueward evolution from the yellow hypergiant stage. Thus, it is possible that the sgB[e] stars may be the hotter counterparts of a class 
of stars that are candidates for  post-red supergiant evolution.  
In Figure 4, we show the spectra of  the two warm hypergiants, M31-004522.52 and M31-004621.08,  that have [Fe II] and [O I] emission with one of 
the B[e] supergiants with the [Ca II] emission, M33C-15731.  

\section{Circumstellar Dust} 

We have emphaisized here and in Paper II, that  the SEDs of the LBVs show free-free emission, but no evidence for warm 
circumstellar dust in the near- or mid-infrared in contrast with both the B[e] supergiants and the warm hypergiants. 
The SEDs for several of these stars belonging  to all three groups are shown in Papers I and II. The Fe II emission line stars, distinct from the B[e] supergiants, also do not have warm dust in their SEDs. To illustrate this difference, we show the SEDs for two of them and two sgB[e] stars from this paper in Figure 5. 
The presence or not of circumstellar dust is included in Table 5 for each of the stars. One of the B[e] 
supergiants, M33-013242.26 does not have warm dust, but PAH emission from  surrounding nebulosity, and M31-004307.1 has free-free 
emission, but weak evidence for circumstellar dust.

\citet{Oksala} and \citet{Kraus} 
demonstrated that the dusty B[e] stars separate from the LBVs in the J$-$K vs. H$-$K two-color diagram. Following their example, in 
Figure 6a we show the near-infrared two-color  diagram for the  LBVs, B[e] supergiants, the Fe II emission line stars, and 
the warm hypergiants. This figure shows more scatter than the two-color figures in the Oksala and Kraus papers. The  
separation between dusty and stars without dust at $H-K$ $\cong$ 0.5 --0.6 mag is not clearcut.  This is partly
because many of the sgB[e] stars are below the magnitude limit for the 2MASS survey, lack JHK magnitudes, and therefore are not 
plotted. Furthermore, stars with strong free-free emission in the near-infrared have larger $H-K$ colors.
The dusty stars are more clearly separated using the longer wavelength IRAC magnitudes, [3.6 -- 4.5] vs [5.8 -- 8.0] shown in Figure 6b. Note that the stars to the lower right in Figure 6b have excess emission in the 8$\mu$m band due to PAH emission from 
surrounding nebulosity.

\section{Other Emission Line Stars}

{\it Hot Supergiants and Yellow Supergiants}

Many of the  stars in Table 2 and most of those in Paper II have emission lines primarily of Hydrogen, many with P Cygni profiles, and some also with He I emission. The question is, are these LBVs or LBV candidates? 
In \citet{RMH2014a} and in Paper II we noted that the outflow velocities of the confirmed LBVs in the quiescent state were significantly lower than their hot supergiant counterparts that they spectroscopically resemble.
Furthermore, the confirmed LBVs and the candidates that resemble hot mass-losing supergiants  also have Fe II and [Fe II] emission lines  in their spectra (Table 5). 
This is of course one of the reasons that the sgB[e] stars are often confused with LBVs. The few LBVs 
that do not have Fe II and [Fe II], such as AF And and Var 2 in M33, are those whose spectra are similar 
to the Of/late WN stars. These spectroscopic characteristics, slow winds and Fe II emission lines,  can be used to help identify potential 
LBV candiates and  separate them  from  the numerous hot supergiants with stellar winds and mass loss  
in the upper  HR diagram.

The same is true for the later type supergiants, those with A-type spectra and the yellow supergiants. During the 
LBV ``eruption'' or dense wind state, their absorption line spectra resemble cooler A to F-type supergiants with apparent temperatures of $\approx 7500 -- 9000\deg$. They not only have strong Hydrogen emission lines with P Cyg profiles, but also Fe II emission lines as seen in the maximum light spectra of Var C \citep{RMH2014a} and J004526.62+415006.3 \citep{RMH15}. In Paper III we identified numerous yellow supergiants (YSGs) with Hydrogen emission line 
spectra many with P Cyg profiles indicative of mass loss.  These  YSGs occupy the same part of the HR diagram as the LBVs in eruption and like the warm hypergiants and the B[e] supergiants can be mistaken for an LBV.  The spectra of 
these stars lacked the Fe II emission and they  were  not LBVs in eruption. Instead we suggested that they were candidates for post-red supergiant evolution.   

Based on the above  characteristics we did not identify any additional LBV candidates among our new spectra descibed in Table 2.  
We also repeated spectroscopy for several hot and intermediate type supergiants discussed in Paper II 
to look 
for potential variability (Table 3). The intermediate type stars or YSGs did not demonstrate any spectral variability and none of them have the Fe II emission lines. The B1Ia + WN supergiant, M33C-9304 did not show any spectral variability.  Three of the hot 
supergiants in M33 however are listed by 
\citet{Massey16} as LBV candidates: M33C-19725(J013339.52+304540.5), B416(J013406.63+304147.8, UIT 301), 
and J013424.78+303306.6. 

We discussed  
J013406.63+304147.8 at length in Paper II. It is a very luminous O9.5 Ia supergiant. Its spectrum 
does indeed have weak 
Fe II and [Fe II] emission, but the only P Cyg profiles are the He I $\lambda$5876 and $\lambda$4471  lines
and the measured outflow velocity of -280 km s$^{-1}$ is somewhat high for an LBV in quiescence. Its spectrum from 2013 did not show any signficant changes in the emission lines. It is also a possible binary. We concluded 
that it was  not an LBV, but a very luminous star or pair of stars. Its characteristics are a mixture, but with the Fe II emission lines it is a possible LBV candidate and deserves further observation.     
We described M33C-19725 (B0.5 I) as peculiar in Paper II due to emission in the core of the He I lines and a second emission component on the red wing of H$\alpha$ and H$\beta$. In its spectrum 
from 2013, the He I lines have changed and now clearly show P Cygni profiles while the P Cyg absorptions have weakened in the Balmer lines suggesting changes in the wind and mass loss rate. The emission feature on the red wing of H$\alpha$ and H$\beta$ is also gone. Its outflow velocity is normal for an early B-type supergiant.  M33C-19725 does not have any 
Fe II or [Fe II] emission lines and based on the LBV characteristics described above,  we do not consider it an LBV candidate. J013424.78+303306.6 is a late B/early A-type supergiant with H emission. It spectrum does not 
have any P Cygni absorption features, however it does show weak Fe II and [Fe II] lines which were noticeably 
weaker in its spectrum  from 2014. Thus it may be  an LBV candidate or a low-luminosity LBV,  possibly in its dense wind state, perhaps 
similar to M31-004425.18 \S {3}.     

\citet{Massey16} included an additional 5 and 8  supergiants  in M31 and M33, respectively,  as possible LBV candidates. We have 
spectra of nine of these stars (Paper II) and obtained spectra of two additional stars from the CfA Optical/Infrared Archive 
for MMT/Hectospec data: J004221.78+410013.4 and J004444.01+415152.0. They are respectively, a B[e] supergiant and an Fe II emission line star and are included in Table 5. Spectra of the other two candidates were not available in the archive.  The spectra of the nine stars\footnote{J004350.50+414611.4, J004507.65+413740.8, J013303.09+303101.8, J013341.28+302237.2, J013344.79+304432.4, J013357.73+301714.2, J013416.44+303120.8, J013422.91+304411.0, J013429.64+303732.1} are described in Paper II, 
Table 1. We examined them again applying the  spectroscopic characteristics for LBV candidates. Except for M33C-4640(J013303.09+303101.8),
none qualify as LBV candidates. They have P Cyg profiles in the Balmer lines which sometimes have broad wings due to stellar winds and mass loss, but no Fe II emission. We discussed M33C-4640 in Paper I and suggested that with its less luminous A-type spectrum
and weak Fe II emission, it might be a post red-supergiant. \citet{BurgPhD} concluded that it did not 
show any significant variability over the preceding 12 years.  Although, it  does not show any evidence for recent mass loss, its spectral characteristics are equally consistent with an LBV candidate of lower luminosity similar to J013424.78+303306.6 discussed above and M31-004425.18.

{\it Of/late-WN stars}\\
Four of the confirmed LBVs, AF And and Var 15 in M31 and Var B  and Var 2 in M33, have spectral characteristics that resemble the Of/late WN type stars. We listed 11 in Paper II and showed  examples of some of their spectra with the spectroscopically similar LBVs. Sixteen more are included here in Table 2. Unfortunately, there is little to spectroscopically 
distinguish the four confirmed LBVs as a group from the larger set Of/late WN stars. This does not mean that all Of/WN stars are candidate LBVs. Of course if they eventually exhibit the 
characteristic  S Dor variability they are LBVs. Short term  spectroscopic changes such as those described in \S {3} for AF And, Var 15, and Var 2 would also be indicators that an Of/late WN star may be a candidate LBV.

In Paper II, we 
found that the Of/WN stars have somwhat higher wind velocities of -329 and  -313 km s$^{-1}$ for the H and He I lines respectvely,  compared to  -229 and -221 km s$^{-1}$ for the four LBVs with Of/late WN type 
spectra.  The spectra of many Of/WN star have P Cygni profiles in their H and He I 
emission lines, and although this is not a large difference\footnote{If we remove the three stars below that have low outflow velocities from the mean calculated in Paper II, then average wind velocities for the Of/late WN stars are -371 and -396 km s$^{-1}$}, we can use the outflow 
velocity measured from the absorption minima to identify potential candidates.  
Nine of the Of/WN stars in Paper II have P Cygni profiles. Three have measured outflow 
velocities that are consistent with the LBVs: M33C-21386(-221 km s$^{-1}$),  
M33C-15235(-233 km s$^{-1}$), and possibly M33-013432.76(-263 km s$^{-1}$).  In this paper, 
ten Of/WN stars have a P Cygni profile, four of which have low outflow velocities; M33C-10788(-255 km s$^{-1}$), M33-013334.06(-224 km s$^{-1}$), M33C-16364(-246 km s$^{-1}$) and 
maybe M33C-5916(-278 km s$^{-1}$), all measured from the He I $\lambda$5876 line.  \citet{Massey16} included three stars with Of/late WN spectra as LBV candidates;  J004341.84+411112.0 has wind velocities of  -337 and -323 km s$^{-1}$ from the H and He I lines, J013332.64+304127.2 (M33C-15742) has  
no P Cyg profiles, and J013351.46+304057.0 is M33C-15235 above.  Thus of the 27 Of/late WN stars for which we have spectra, 7 have relatively low outflow velocities and may be considered LBV candidates.   

{\it The PSO stars}\\
We also obtained spectra of four LBV candidates in M31 proposed by \citet{Lee} based on variability in the 
Pan-STARRS survey. 
 All four showed variability in the visual on the order of 0.4 to 0.5 mag and had H-K colors $\leq$ 0.5 mag like the confirmed LBVs, see Fig 6a this paper. 
Two of the stars were fainter than 19th mag and consequently have poor S/N spectra that did permit a spectral classification, although the red spectrum of PSO-J10.8180+41.6265 shows molecular bands and thus 
has a blended spectrum. The other two (PSO-J11.0457+41.5548 and PSO-J11.2574+42.0498) have 
absorption line spectra of typical of 
B and and A-type supergiants with H and He I emission lines, but no P Cyg profiles and no Fe II emission lines. 
The red spectrum of PSO-J11.2574+42.0498  also has molecular bands in the red and the red spectrum  
of PSO-J11.0457+41.5548 has [O I] emission  which we attribute to the strong nebular contamination in 
its red spectrum. Despite their small H-K colors, both PSO-J10.8180+41.6265 and PSO-J11.2574+42.0498  
show mid-infrared excess emission in the IRAC bands probably due to their likely M star companions.
At this time there is insufficient information to include them as LBV candidates.

\section{Summary:  LBVs vs. B[e]supergiants vs.Warm Hypergiants}

The numerous luminous stars with emission lines in the upper HR Diagram share  spectroscopic properties which at first glance may make them difficult to separate. In this paper we have taken a 
closer look at their spectra and SEDs, and conclude that it is relatively straightforward to identify and separate the LBVs, the B[e] supergiants and warm hypergiants:  

\begin{itemize}  
\item{B[e] supergiants have emission lines of [O I] and [Fe II] and some also have [Ca II] in their spectra. Most of the spectroscopically confirmed sgB[e] stars also have warm circumstellar dust in their SEDs} 
\item{Confirmed LBVs do not have the [O I] emission lines in their spectra. Some LBVs have [Fe II] emission lines, but not all. Their SEDs show free-free emission  in the near-infrared but {\it no warm dust}. Their most important and defining characteristic is the S Dor-type variability.}
\item{The warm hypergiants spectroscopically resemble the LBVs in their eruption or dense wind state and the B[e] supergiants. However, they are very dusty. Some have [Fe II] and [O I] emission in their spectra 
like the sgB[e] stars, but can be distinguished from them by 
their absorption line spectra characteristic of A and F-type supergiants. In contrast, the B[e] supergiants have strong continua and few if any apparent absorption lines.}
\end{itemize}

Candidate LBVs should share the spectral characteristics of the confirmed LBVs plus the lack of warm 
circumstellar dust. These include P Cygni profiles in the H and/or He I lines for those with 
 spectra like the hot supergiants.  Fe II emission is present, but not [O I]. [Fe II] emission may also be present. Some confirmed LBVs resemble the Of/late WN stars,  and although there is no clearcut 
spectral difference, a low outflow velocity, $\le$ 300 km s$^{-1}$, will help to identify potential candidates among the Of/late WN stars. Based on these criteria the Fe II 
emission line stars (no [O I] and no dust, see \S {4}, Table 5)), should be considered LBV 
candidates. Table 7  is a list of candidate LBVs in M31 and M33 in our data based on the guidelines and
discussion presented here. 

The spectroscopic differences and other characteristics, such as the presence or lack of
circumstellar dust, may be clues to other parameters that are more fundamental to the
astrophysics of these stars, such as rotation, binarity and evolutionary state. For example, the warm
hypergiants are likely post-RSGs. They may transition to B[e] supergiants or even to the
less luminous LBVs once they shed their dust. And as is often the case in astronomy, there
will be some stars that do not quite fit our description or may be  transition objects such 
as B526NE.

\acknowledgements
Research by R. Humphreys on massive stars is supported by  
the National Science Foundation AST-1109394.  J. C . Martin's collaborative work 
on luminous variables is supported by the National Science Foundation grant  AST-1108890.  In addition to our own data observed with the MMT/Hectospec, we use spectra obtained from  the CFA Optical/Infrared Science Archive .  This paper uses data from the MODS1 spectrograph built with funding from NSF grant AST-9987045 and the NSF Telescope System Instrumentation Program (TSIP), with additional funds from the Ohio Board of Regents and the Ohio State University Office of Research.  This publication also makes use of data products from the Wide-field Infrared Survey Explorer, which is a joint project of the University of California, Los Angeles, and 
the Jet Propulsion Laboratory/California Institute of Technology, funded by the 
National Aeronautics and Space Administration.

{\it Facilities:} \facility{MMT/Hectospec, LBT/MODS1}


\begin{deluxetable}{llcc}
\tablecaption{Journal of Observations\label{tab:one}}
\tabletypesize{\footnotesize}
\tablecolumns{4}
\tablewidth{0pt}
\tablenum{1}
\tablehead{
\colhead{Target}    & 
\colhead{Date}      &
\colhead{Exp. Time} &
\colhead{Grating, Tilt} \\
\colhead{}          &
\colhead{(UT)}      &
\colhead{(minutes)} &
\colhead{}
  }
  \startdata
  M31A-Blue & 2013 Sep 25 & 120 & 600l, 4800\AA \\
  M31A-Red & 2013 Sep 26 & 90 & 600l, 6800\AA \\
  M31B-Blue & 2013 Oct 12 & 120 & 600l, 4800\AA \\
  M31B-Red & 2013 Oct 9 & 90 & 600l, 6800\AA \\
  M33-Blue & 2013 Oct 7 & 120 & 600l, 4800\AA \\
  M33-Red & 2013 Oct 7 & 90 & 600l, 6800\AA \\
  M33-Blue & 2014 Nov 29 & 120 & 600l, 4800\AA \\
  M33-Red & 2014 Nov 16 & 90 & 600l, 6800\AA \\
  M31A-Blue & 2015 Sep 20 & 120 & 600l, 4800\AA \\
  M31A-Red & 2015 Sep 20 & 90 & 600l, 6800\AA \\
  M31B-Blue & 2015 Sep 20 & 120 & 600l, 4800\AA \\
  
  \enddata
\end{deluxetable}

\begin{deluxetable}{lcccl}
\rotate
\tablewidth{0 pt}
\tabletypesize{\footnotesize}
\tablenum{2} 
\tablecaption{Luminous Stars and Variables in M31 and M33: New Spectra 2013 -- 2015 }
\tablehead{
\colhead{Star Name} &
\colhead{RA and Dec (2000)} &
\colhead{Yr}   & 
\colhead{Spec. Group} &
\colhead{Other Id/Notes/References}  
}
\startdata 
   &   &  M31   &    &    \\
  M31-003910.85 &  00:39:10.84	40:36:22.3 &  13  &  Fe II Em. Line    & H em, br.wings, He I P Cyg, Fe II em \\
  M31-003944.71 &  00:39:44.70	40:20:56.1  &  13 &  H II             & strong neb em, He I abs. \\   
 PSO-J10.1165+40.7082 &  00:40:27.99 40:42:29.0 & 15 &    \nodata      &  low S/N, H$\beta$  \\
  M31-004030.28 &  00:40:30.27	40:42:33.0  &  13 &  Hot Supergiant   &  B1-B2 I, H em, P Cyg, neb em \\
  M31-004032.37 &   00:40:32.36	40:38:59.7  &  13 &  Hot Supergiant   &  B5 I, neb em  \\
  M31-004033.80 &   00:40:33.79	40:57:17.1  &  13 &  Hot Supergiant   &  B0 I, H em, P Cyg, H$\alpha$ br. wings, neb em \\  
  M31-004043.10 &   00:40:43.09	41:08:45.9  &  13 &  Fe II Em. Line(sgB[e])   & H em br wings,[Fe II] em  \\
  M31-004051.59 &   00:40:51.58	40:33:02.9  &  13,15 &  LBVc/Intermed Type &  early A-type, see \S {3.1}, (Paper III) \\  
  M31-004052.19 &   00:40:52.18	40:31:16.5  &  13 &  Intermed Type   &  A0 I         \\
  M31-004056.49 &   00:40:56.48	41:03:08.6  &  13 &  Of/late-WN       &  H em, br.wings, He I em, He II4686, (Ofpe/WN9 \citet{Massey07})\\ 
\enddata
\tablenotetext{a}{LBT/MODS spectrum, Nov. 2014, see text}
\end{deluxetable}

\begin{deluxetable}{lcccl}
\rotate
\tablewidth{0 pt}
\tabletypesize{\footnotesize}
\tablenum{3} 
\tablecaption{Repeat Spectroscopy}
\tablehead{
\colhead{Star Name} &
\colhead{RA and Dec (2000)}  &
\colhead{Yr}  &  
\colhead{Spec. Group} &
\colhead{Other Id/Notes/References}  
}
\startdata 
   &    &     M31   &    &    \\
   &  J004247.30  +414451.0 &  10,15 & Intermed-Type & F2 Ia (Paper II)\\
AE And  &      J004302.52   +414912.4 & 10,13,15 &  LBV  & see \S {3}, (Paper II)  \\
   &  J004322.50  +413940.9 & 10,13,15  & Warm Hypergiant  & late A --F0 I (Paper I)  \\
AF And  &   J004333.09  +411210.4  & 10,13,15 & LBV   &  see \S {3}, (Paper II)  \\
   &  J004337.16  +412151.0 & 10,13  & Intermed-Type & F8 I (Paper II) \\ 
Var 15  &       004419.43   +412247.0 & 10,13,15 &  LBV          &  see \S {3}, (Paper II) \\
   &   J004424.21  +412116.0  & 10,15  &  Intermed-Type & F5 Ia, (Paper II)  \\
   &   J004425.18\tablenotemark{b}    +413452.2 & 12,13,15  &  LBV candidate &  B0-1 Ia, see \S {3.1}, (Paper II)  \\
   &   J004444.52 +412804.0   &  10,12,13,15 &  Warm Hypergiant & F0 Ia (Paper I) \\
Var A-1  &  J004450.54  +413037.7 & 10,13,15  & LBV   &  see \S {3}, (Paper II)\\ 
   &   J004522.58  +415034.8 & 10,12,13,15    &  Warm Hypergiant & A2 Ia (Paper I) \\ 
   &   J004526.62  +415006.3 & 10,13,15  &  LBV &  see \S {3}, \citet{RMH15} \\ 
   &               &              &             &                  \\
   &               &   M33   &         &                                \\
Var A & J013232.80 +303025.0  & 10,11,13,14  &  Warm Hypergiant & F8 Ia (Paper I) \\
M33C-14239   &  J013248.23  +303950.3  & 10,14 & Fe II Em. Line    &  He I em, H$\alpha$ br wings,  Fe II, [FeII] em, (Paper II)  \\
M33C-4119  & J013312.81  +303012.7 & 10,11,14  &  Hot Supergiant & B5 Ia, H, He I em., P Cyg (Paper II)\\ 
          &  J013324.62  +302328.4 & 10,14 & Fe II Em. Line(sgB[e])  & [Fe II] em., asymmetric H$\alpha$, (Paper II)   \\ 
N045901  &  J013327.40  +303029.4 & 10,12,14  & Intermed-Type  & F: pec I, (Paper II) \\ 
M33C-7256 &  J013333.22  +303343.4 & 10,13,14 & Fe II Em. Line(sgB[e])   & [Fe II], He I em., neb. em. \\ 
Var C     &      J013335.14 +303600.4  &  10,13 &  LBV        &  see \S {3}, Paper II, \citet{RMH2014a}  \\ 
M33C-19725 & J013339.52  +304540.5  &  10,13 & Hot Supergiant   & B0.5: I pec, B517 \citet{HS80}, Paper II \\  
Var B     &     J013349.23  +303809.1  &  10,13,14 & LBV     &  see \S {3}, Paper II  \\ 
M33C-15731 &     J013350.12  +304126.6 &  10,12,13,14 &  Fe II Em. Line(sgB[e])  & Ca II, [Ca II], [Fe II], wk He I em., H P Cyg,  UIT212, (Paper II)  \\  
N093351  &     J013352.42 +303909.6 & 10,11,13  &  Warm Hypergiant & F0 Ia (Paper I), M33C-13568  \\  
B324  &  J013355.96 +304530.6 & 10,11,13,14 &Warm Hypergiant  &  A8-F0 Ia, UIT 247, (Paper I) \\
M33C-9304  & J013358.70 +303526.5  &  10,13 &  Hot Supergiant & B0-B1 Ia + WN, H P Cyg, asymmetric (B1Ia+WNE, UIT267), (Paper II)  \\ 
  &       J013406.63    +304147.8 &  10,13  & Hot Supergiant & O9.5 Ia, B416 \citet{HS80}, UIT301, (Paper II)  \\ 
Var 83  &       J013410.93  +303437.6 & 10,13,14 &  LBV   &  see \S {3}, (Paper II)        \\ 
N125093 &  J013415.38 +302816.3   &  10,12,13      &  Warm Hypergiant &  F0 - F2 Ia, Paper I \\
Var 2   &      J013418.36  +303836.9 & 10,13,14 &  LBV   &  see \S {3}, (Paper II)       \\
        &   J013424.78  +303306.6 &  10,14 &  Hot Supergiant & B8-A0 Ia, H$\alpha$ asymmetric, (Paper II) \\ 
	&  J013459.47 +303701.9 & 10,14  & Fe II Em. Line(sgB[e])     & weak [Fe II], He I em.  (Paper II)   \\
        &   J013500.30 +304150.9 & 10,14  & Fe II Em. Line(sgB[e])     &  [Fe II], (Paper II)     \\ 
GR 290   &  J013509.73   +304157.3 &  10,13 & Of/late-WN & M33-V532, Romano's star, (Paper II) \\
\enddata
\end{deluxetable}

\begin{deluxetable}{llllllllllllllllll}
\rotate
\tablewidth{0 pt}
\tabletypesize{\scriptsize}
\tablenum{4} 
\tablecaption{Multi-Wavelength Photometry }
\tablehead{
\colhead{Star} & 
\colhead{U}  &
\colhead{B} &
\colhead{V} &
\colhead{R} & 
\colhead{I} &
\colhead{J} &
\colhead{H} &
\colhead{K} &
\colhead{3.6$\mu$m}\tablenotemark{a} &
\colhead{4.5$\mu$m}\tablenotemark{a} &
\colhead{5.8$\mu$m}\tablenotemark{a} &
\colhead{8$\mu$m}\tablenotemark{a}  &
\colhead{3.4$\mu$m}\tablenotemark{b} &
\colhead{4.6$\mu$m}\tablenotemark{b} &
\colhead{12$\mu$m}\tablenotemark{b} &
\colhead{22$\mu$m}\tablenotemark{b} &  
\colhead{Var}
}

\startdata
   &    &   &    &  & & & & M31   & &   &  &  &  &   & &   \\ 

M31-003910.85 & 17.42 & 18.46 & 18.18 & 17.81 & 17.54 & \nodata & \nodata &\nodata & \nodata & \nodata &\nodata &\nodata & \nodata & \nodata &\nodata &\nodata \\
M31-003944.71 & 17.44 & 18.35 & 18.20 & 18.07 & 17.95 & \nodata & \nodata &\nodata & \nodata & \nodata &\nodata &\nodata & \nodata & \nodata &\nodata &\nodata \\
PSO-J10.1165+40.7082  & 18.50 & 19.54 & 19.51 & 18.94 & 17.55 & 15.62 & 14.73 & 14.32 & 13.58 & 13.19 & 12.88 & 12.03 & 13.82 & 13.56 & 10.37 & 7.92 & var.\\ 
M31-004030.28 & 16.18 & 17.31 & 17.36 & 17.31 & 17.26 & 16.72 & 15.63 & 15.38 & \nodata & \nodata &\nodata &\nodata & \nodata & \nodata &\nodata &\nodata \\ 
M31-004032.37 & 17.49 & 18.20 & 17.76 & 17.45 & 17.09 & 16.32 & 15.69 & 15.17 & 15.18 & 14.90 & 11.97 & 10.16 & \nodata & \nodata &\nodata &\nodata \\ 
M31-004033.80 & 16.42 & 17.36 & 17.33 & 17.23 & 17.25 & \nodata & \nodata &\nodata & \nodata & \nodata &\nodata &\nodata & 14.88 & 14.92 & 10.77 & 8.65\\
M31-004043.10 & 17.70 & 18.77 & 18.62 & 17.56 & 17.72 & 17.44 & 15.82 & 15.30 & 14.90 & 14.60 & 13.75 & 13.38 & 13.56 & 12.58 & 10.50 & 8.55\\
M31-004051.59 & 16.44 & 17.21 & 16.99 & 16.77 & 16.58 & 16.38 & 15.59 & 15.75 & \nodata & \nodata &\nodata &\nodata & 14.43 & 12.30 & 9.92 & 8.21\\
M31-004052.19 & 17.28 & 18.02 & 17.69 & 17.48 & 17.25 & \nodata & \nodata &\nodata & \nodata & \nodata &\nodata &\nodata & \nodata & \nodata &\nodata &\nodata \\
M31-004056.49 & 17.01 & 18.00 & 18.09 & 18.00 & 17.89 & \nodata & \nodata &\nodata & \nodata & \nodata &\nodata &\nodata & \nodata & \nodata &\nodata &\nodata \\
\enddata
\tablenotetext{a}{{\it Spitzer}/IRAC}
\tablenotetext{b}{WISE}
\end{deluxetable}

\begin{deluxetable}{lccccccccl}
\rotate
\tablewidth{0 pt}
\tabletypesize{\footnotesize}
\tablenum{5} 
\tablecaption{Emission Line Clues in the Fe II Emission-Line Stars and  LBVs}
\tablehead{
\colhead{Star Name} &
\colhead{He I} &
\colhead{Fe II}  &
\colhead{[Fe II]}  &  
\colhead{[O I]} &
\colhead{[Ca II]} &
\colhead{Ca II}  &
\colhead{O I} &
\colhead{CS Dust} &
\colhead{Remark} 
}
\startdata 
\sidehead{B[e] and Fe II Emission Line Stars}
   &    &   &   &  M31   &    &  &  &  & \\
M31-003910.85 &  yes &  yes &  no &  no  &  no  &  \nodata & no & \nodata &  \\
M31-004043.10 &  yes &  yes &  yes & yes &  yes &  \nodata & yes & ?(f-f) & sgB[e] \\
M31-004057.03 &  ?   &  yes &  yes &  yes &  ?  &  \nodata &  ?  &  \nodata & sgB[e] \\
M31-004220.31 &  ?   &  yes &  yes &  yes &  no &   \nodata & abs & yes & sgB[e] \\
M31-004221.78 &  yes & yes &   yes &  yes &  no &  yes      & yes &  yes  & sgB[e], \citet{Massey16} \\
M31-004229.87 &  yes &  yes &  yes &  yes &  no &  yes      & yes & yes &  sgB[e], Paper II \\
M31-004320.97 &  yes &  yes &  yes &  yes &  yes &   \nodata & yes &  yes & sgB[e] Paper II \\
M31-004411.36 &  yes &  yes &  no  &  ?   &  yes &   \nodata & yes &  no (PAH) &  Paper II  \\
M31-004415.00 &  yes &  yes &  yes &  yes &  yes &   \nodata & yes &  yes & sgB[e], Paper II \\
M31-004417.10 &  yes &  yes &  yes &  yes &  yes &   yes  & yes &  yes & sgB[e], Paper II \\
M31-004442.28 &  yes &  yes &  yes &  yes &  yes &   \nodata & yes & yes & sgB[e], Paper II \\  
M31-004444.01 &  yes &  yes &  no  &  no  &  no  &   yes     &  no  & no  &  \citet{Massey16} \\ 
   &    &   &   &  M33   &    &  &  &   \\ 
M33C-2976   &   no  &   yes  &  no  &  no  &  yes? &   \nodata &  no &  no &     \\
M33C-4174   &   yes &  yes   &  no  &  no  &  no   &   \nodata &  yes &  no  & Paper II\\
M33-013242.26 & yes &  yes   & yes  &  yes &  ?    &   \nodata &  yes &  no (PAH)  & sgB[e], Paper II   \\ 
M33C-14239  &  yes  &   yes  &  no  &  no  &  no   &   \nodata &  yes &  no (PAH)  & Paper II \\ 
M33C-25255  &  yes  &   yes  &  no  &  no & no   & \nodata &  no  &  no  &    \\
M33-013317.22 &  no & yes    &  no  &  no & no   &  \nodata &  no &  no (f-f) &    \\ 
M33-013324.62 &  yes &  yes & yes  &  yes  &  no  &  \nodata &  no & yes & sgB[e], Paper II\\
M33C-7256  &  yes  &  yes  &  yes  &  yes  &  no  &  \nodata &  yes &  yes & sgB[e], Paper II \\
M33C-7024  &  yes &  yes &  no &  no &  no &    \nodata &  no &  \nodata &     \\
M33C-6448  &  yes &  yes &  yes &  yes & yes &    \nodata &  yes & \nodata &  sgB[e] \\   
M33C-24812 &  no  &  yes &  yes  &  yes &  no &  \nodata &  no & yes &  sgB[e]    \\
M33C-15731 &  yes & yes &  yes &  yes & yes  &  yes  &  yes &  yes &  sgB[e], Paper II\\   
M33-013426.11 &  yes & yes & yes & yes & yes &  \nodata &  no & yes & sgB[e], Paper II\\  
M33C-20109 &  yes &  yes &  ?    &  no  &  yes & \nodata &  no &   no (f-f) &   \\
M33-013459.47 & yes & yes & yes & yes &  no  & \nodata &  yes & yes  &   sgB[e], Paper II\\
M33-013500.30 &  no & yes & yes & yes &  no  & \nodata &  no  &  yes  & sgB[e], Paper II\\ 
\sidehead{LBVs} 
AE And &   yes &  yes &  yes &  no  &  no &  \nodata  &  abs & no  &  \S {3}, Paper II\\
AF And &   yes &  no(Fe III)  &  no  &  no  &  no & \nodata & yes P Cyg & no  &   \S {3}, Paper II\\
Var 15 &   yes & yes &  yes   &  no &  no &  \nodata & yes P Cyg &   no  &  \S {3}, Paper II\\
       &   no  & yes  &   no   &  no &  no &  \nodata & yes P Cyg &  no  &  \S {3}, 2015 sp. \\
Var A-1 &  yes & yes &   yes  &  no &  no &  \nodata & yes P Cyg &  no  &  \S {3}, Paper II \\   
M31-004526.62\tablenotemark{a}  &  yes &  yes &  yes: & \nodata & \nodata & \nodata &  \nodata & no & \citet{RMH15} \\
Var 83  &  yes & yes &   yes &  no &  no &  \nodata & yes P Cyg &  no  &  \S {3}, Paper II \\ 
Var B   &  yes & no  &  no([Fe III])  &  no &  no & \nodata & yes P Cyg &  no  & \S {3}, Fe II \& [Fe II] 2004, Paper II \\
Var 2   &  yes & no  &  no      &  no  &  no  & \nodata & yes  &  no & \S {3}, Paper II \\
Var C   &  yes & yes &  yes     &  no  &  no  & \nodata & abs  &  no  & \citet{RMH2014a}  \\
\sidehead{Candidates}
M31-004425.18 &  yes &  no  & no &  no  &  no  & no  & abs & no  & \S {3.1}, Paper II \\ 
M31-004051.59 &  no  & yes  & no &  no  &  no  & \nodata &  abs &  no  & \S {3.1}, Paper III  \\  
UIT 008       &  yes &  no  & no &  no  &  no  & \nodata &  no  &  no  & {3.1}, Paper II \\
B526NE        &  no  & yes  &  yes & yes  &  no  &  no     &  yes & no  & \S {3.1} \\ 
\enddata
\tablenotetext{a}{The spectrum from 2006 when it was in its quiescent state \citet{RMH15}. There was no red spectrum.} 
\end{deluxetable}

\begin{deluxetable}{lcccccccl}
\rotate
\tablewidth{0 pt}
\tabletypesize{\footnotesize}
\tablenum{6} 
\tablecaption{Emission Line Clues in the Warm Hypergiants}
\tablehead{
\colhead{Star Name} &
\colhead{He I} &
\colhead{Fe II}  &
\colhead{[Fe II]}  &  
\colhead{[O I]} &
\colhead{[Ca II]} &
\colhead{Ca II}  &
\colhead{O I} &
\colhead{Remark} 
}
\startdata 
M31-004322.50 &  no & yes & no &  weak &  yes & \nodata  &  abs &  Paper I \\
M31-004444.52 &  no & yes & no &  weak &  yes &  yes     &  abs &   Paper I \\
M31-004522.58 &  no & yes & yes & yes  &  yes &  yes     &  abs &   Paper I \\
M31-004621.08 &  no & yes & yes & yes &  yes &  yes     &  abs &  Paper III \\    
M33 Var A     &  no & yes & no  & weak & yes &  yes     &  abs &   Paper I \\
M33 B324      &  no & yes & no  &  no  & yes &  yes     &  abs &   Paper I \\
N093351       &  no & no  & no  & weak & yes &  yes     &  abs &   Paper I \\
N125093       &  no & no  & no  &  weak & yes &  yes     &  abs &   Paper I \\
\enddata
\end{deluxetable}

\begin{deluxetable}{lccl}  
\tablewidth{0 pt}
\tabletypesize{\footnotesize}
\tablenum{7} 
\tablecaption{LBV Candidates in M31 and M33}
\tablehead{
\colhead{Star Name} &
\colhead{RA and Dec (2000)}  &
\colhead{Spec. Group} &
\colhead{Notes}  
}
\startdata 
    &    M31  &     &    \\
M31-003910.85 &  00:39:10.84 +40:36:22.3  & Fe II Em line &     \\
M31-004051.59 &  00:40:51.58 +40:33:02.9 &  Intermed Type   & post-RSG?,  see \S {3.1} \\
              &  00:44:11.35 +41:32:57.1 &  Fe II Em  line  &  Paper II \\  
M31-004425.18 &  00:44:25.17 +41:34:52.1  &  Hot Supergiant & probable LBV, see \S {3.1} \\
              &  00:44:44.00 +41:51:51.8  & Fe II Em. line  &         \\ 
    &    M33  &     &    \\ 
M33C-2976    &  01:32:29.0 +30:28:19.5 &  Fe II Em. Line   &      \\  
M33C-4174    &  01:32:35.22+30:30:17.5 &  Fe II Em. Line   &      \\ 
UIT 008       &  01:32:45.38 +30:38:58.2  & Of/late WN  &  Paper II, see \S {3.1} \\
M33C-14239    &  01:32:48.23 +30:39:50.3  &  Fe II Em. Line   &  Paper II \\
M33C-4640     &  01:33:03.10 +30:31:01.9  &  Intermed Type & post-RSG?, Paper II, see \S {7}\\ 
M33C-25255   &  01:33:17.02 +30:53:29.8  &  Fe II Em. Line  &     \\ 
M33-013317.22 & 01:33:17.19 +30:32:01.5  &  Fe II Em. Line  &     \\
M33-013334.06 & 01:33:34.03  +30:47:44.2  &  Of/late-WN  &      \\ 
M33C-7024    & 01:33:37.32  +30:33:28.9  &  Fe II Em. Line  &     \\
M33C-15235   & 01:33:51.43  +30:40:56.9  &  Fe II Em. Line  &  Paper II\\ 
M33C-5916    & 01:33:54.82  +30:32:22.7  &  Of/late-WN &     \\
UIT 301/B416 & 01:34:06.63 +30:41:47.7   &  Hot Supergiant & luminous O star, Paper II, see \S {7} \\
M33C-21386   & 01:34:06.77 +30:47:26.9  & Of/late WN  &  Paper II \\
M33C-10788   & 01:34:16.04  +30:36:42.0   &  Of/late-WN  &  \\ 
             & 01:34:24.78 +30:33:06.6    &  Hot Supergiant & Paper II, see \S {7}\\
M33C-20109   &  01:34:27.08  +30:45:59.7 &  Fe II Em  Line  &     \\  
             & 01:34:32.73 +30:47:17.1   & Of/late WN  &  Paper II \\
M33C-16364   &  01:34:59.36 +30:42:01.1 &  Of/late-WN  &         \\ 
\enddata
\end{deluxetable}


\begin{figure}
\figurenum{1}
\epsscale{0.7}
\plotone{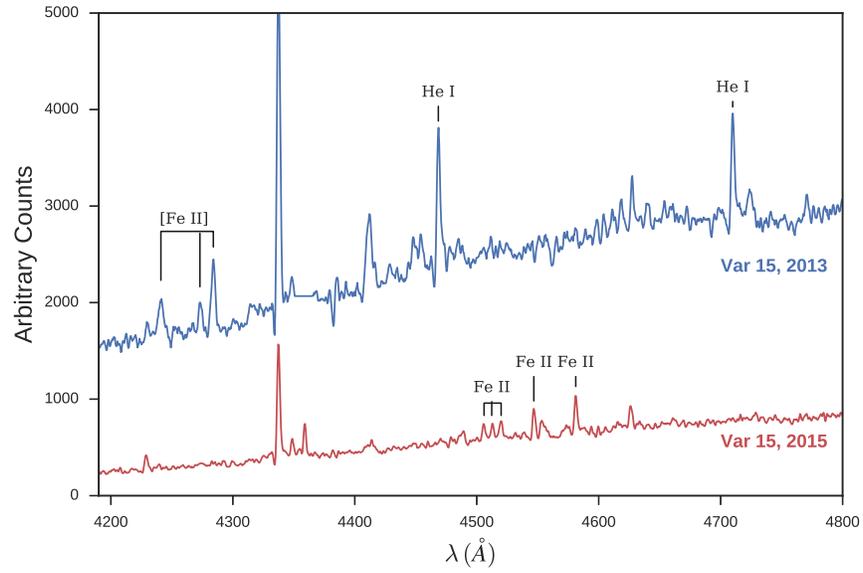}
\caption{A comparison of the 2013 and 2015 spectra of Var 15 in M31 illustrating the disappearance of the He I and [Fe II] emission lines and the weakening of the N II emission.}
\end{figure}

\begin{figure}
\figurenum{2}
\epsscale{0.5}
\plotone{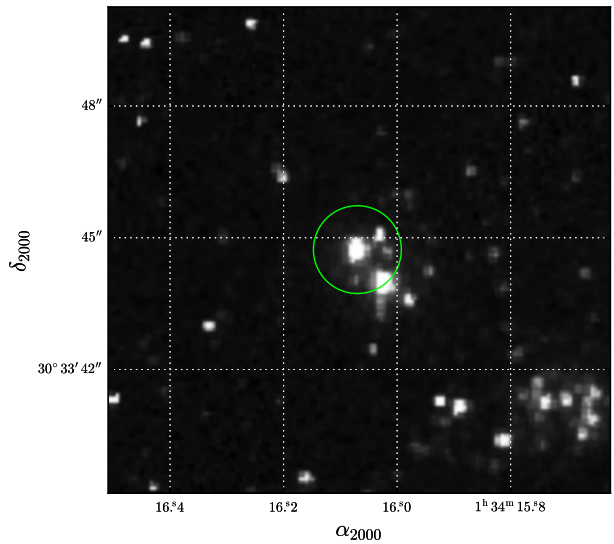}
\caption{HST/WFPC2 F336W image from 2007 (HST ID 8059, PI Casertano). The green circle has a radius of 1$\arcsec$. The star at the center is B526NE, the candidate LBV.}  
\end{figure}

\begin{figure}
\figurenum{3a}
\epsscale{0.7}
\plotone{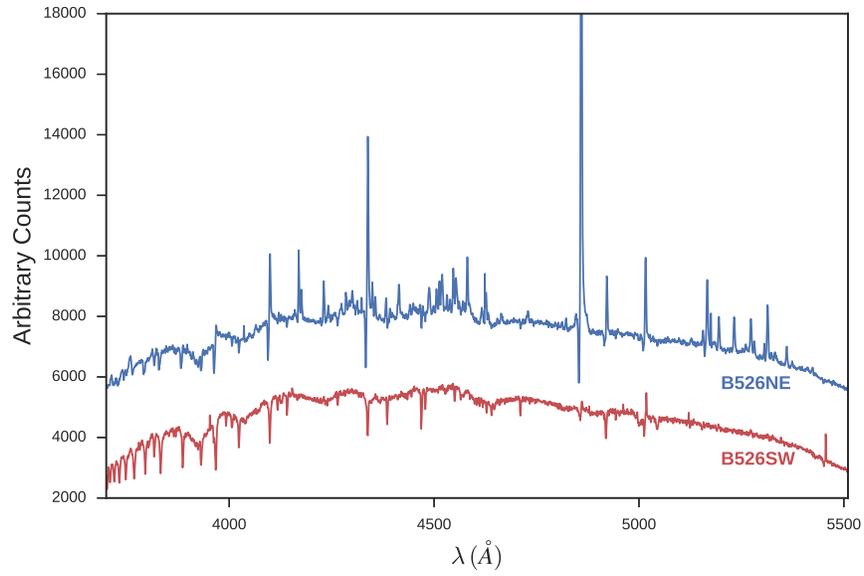}
\caption{The blue spectra of B526NE and B536SW  illustrating the strong Fe II emission libne spectrum and P Cygni profiles in B526NE. These spectra were observed with the LBT/MODS1 spectrograph.}  
\end{figure}

\begin{figure}
\figurenum{3b}
\epsscale{0.7}
\plotone{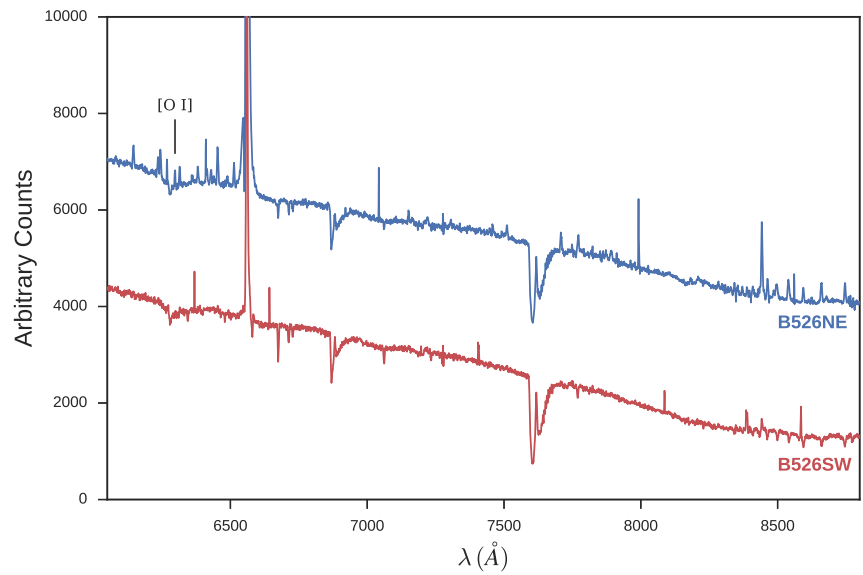}
\caption{The red spectra of B526NE and B526SW. Note the [O I] emission line in B526NE.}
\end{figure}

\begin{figure}
\figurenum{4a}  
\epsscale{0.7}
\plotone{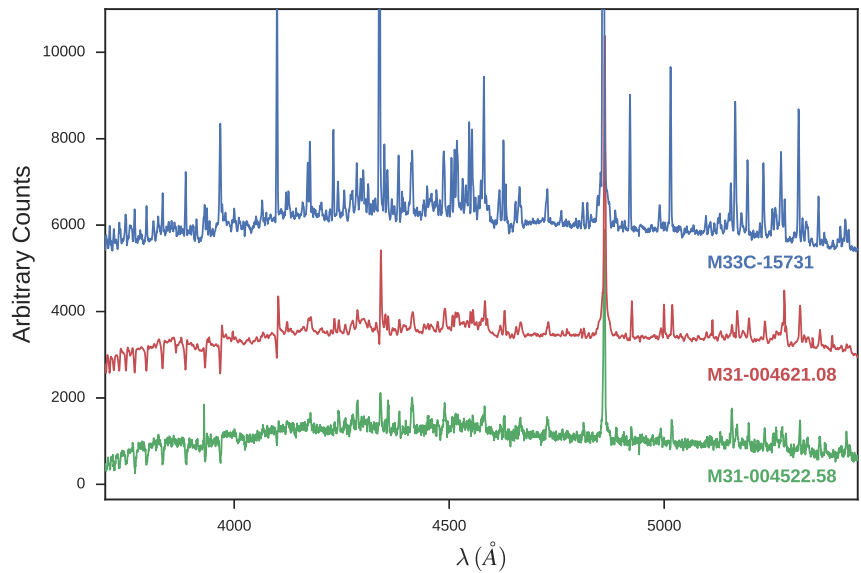}
\caption{The blue spectra of the hypergiants M31-004522.58 and M31-004621.08 and the B[e] supergiant M33C-15731. Although their emission line spectra are similar the hypergiants, have strong absorption in the Balmer lines and the Ca II K line similar to an early A-type supergiant.}
\end{figure}

\begin{figure}
\figurenum{4b}
\epsscale{0.7}
\plotone{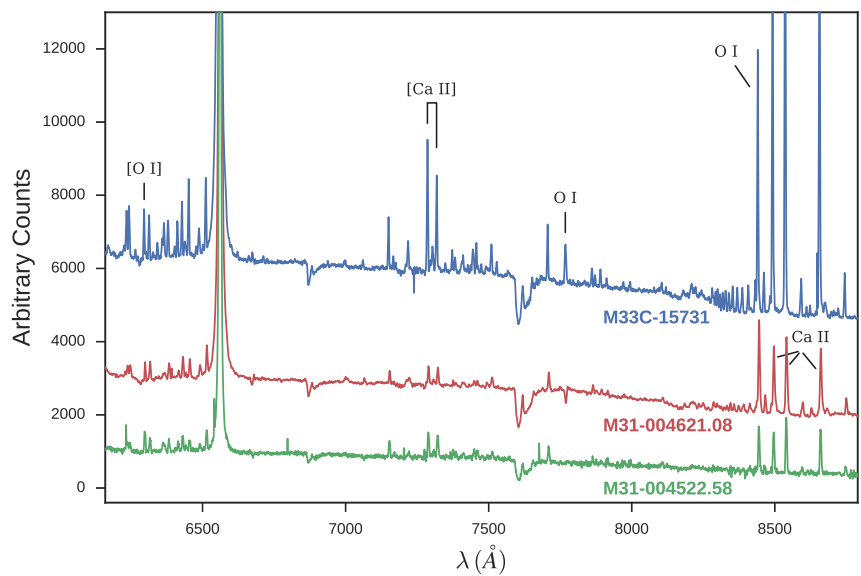}
\caption{The red  spectra of the hypergiants M31-004522.58 and M31-004621.08 with [O I] emission and the B[e] supergiant M33C-15731 which has the [Ca II] and Ca II emission lines like the hypergiants. }
\end{figure}

\begin{figure}
\figurenum{5}
\epsscale{0.8}
\plotone{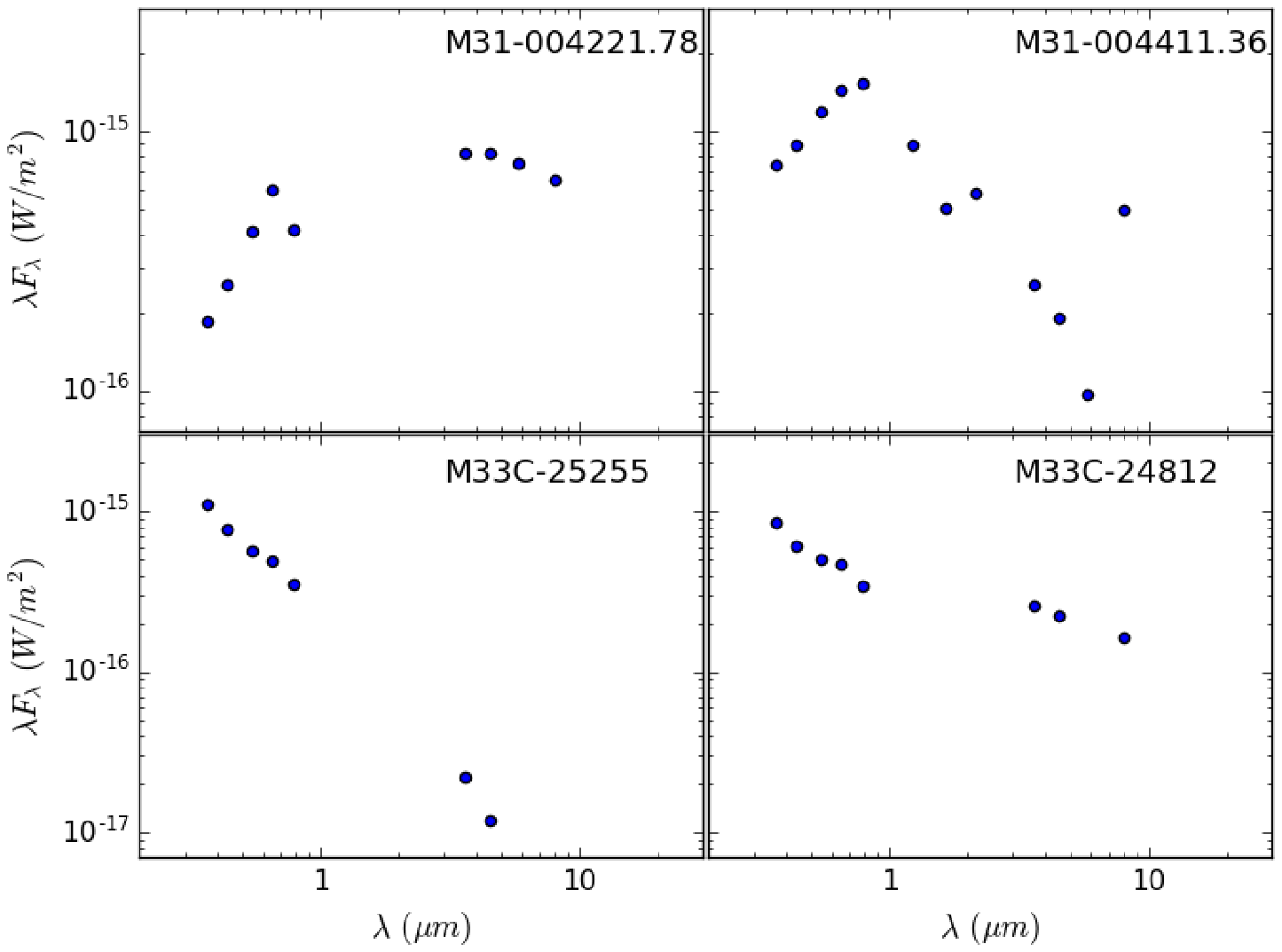}
\caption{Sample SEDs for B[e] supergiants and Fe II emission line stars. The B[e] supergiants, M31-004221.78 and M33C-24812, have circumstellar dust in the 3 to 8$\mu$m region while the two Fe II emission line stars do not. The 8$\mu$m point in M31-004411.36 is due to PAH emission.}
\end{figure}

\begin{figure}  
\figurenum{6a}
\epsscale{0.7}
\plotone{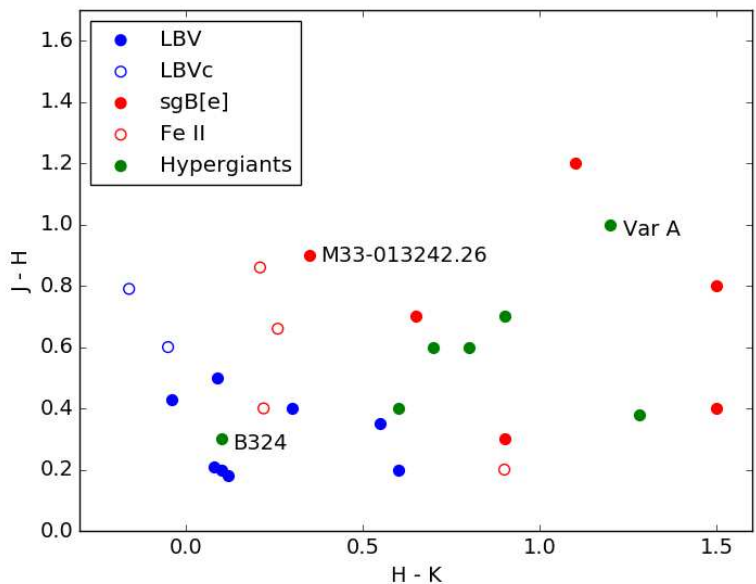}
\caption{The JHK two-color diagram. The hypergiant B324 and the sgB[e] M333-013242.26 do not have circumstellar dust. }
\end{figure}

\begin{figure}
\figurenum{6b}
\epsscale{0.7}
\plotone{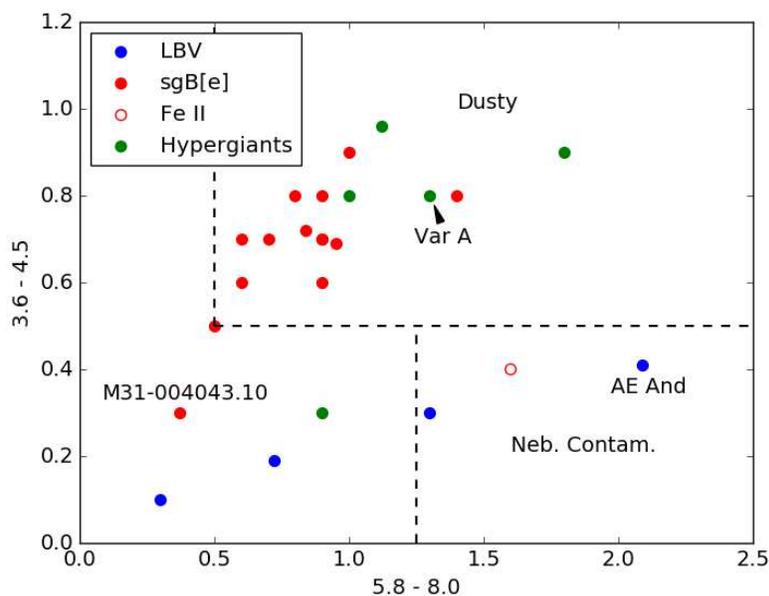}
\caption{The [3.6 -- 4.5$\mu$m] vs [5.8 -- 8.0$\mu$m] diagram. The sgB[e] M31-004043.10 has free-free 
emission but weak evidence for dust. The stars in the lower right have excess emission in the 8$\mu$m band due to nebular emission.}
\end{figure}

\begin{deluxetable}{lcccl}
\rotate
\tablewidth{0 pt}
\tabletypesize{\footnotesize}
\tablenum{2-online} 
\tablecaption{Luminous Stars and Variables in M31 and M33: New Spectra 2013 -- 2015 }
\tablehead{
\colhead{Star Name} &
\colhead{RA and Dec (2000)} &
\colhead{Yr}   & 
\colhead{Spec. Group} &
\colhead{Other Id/Notes/References}  
}
\startdata 
   &   &  M31   &    &    \\
  M31-003910.85 &  00:39:10.84	40:36:22.3 &  13  &  Fe II Em. Line    & H em, br.wings, He I P Cyg, Fe II em \\
  M31-003944.71 &  00:39:44.70	40:20:56.1  &  13 &  H II             & strong neb em, He I abs. \\   
 PSO-J10.1165+40.7082 &  00:40:27.99 40:42:29.0 & 15 &    \nodata      &  low S/N, H$\beta$ em \\
  M31-004030.28 &  00:40:30.27	40:42:33.0  &  13 &  Hot Supergiant   &  B1-B2 I, H em, P Cyg, neb em \\
  M31-004032.37 &   00:40:32.36	40:38:59.7  &  13 &  Hot Supergiant   &  B5 I, neb em  \\
  M31-004033.80 &   00:40:33.79	40:57:17.1  &  13 &  Hot Supergiant   &  B0 I, H em, P Cyg, H$\alpha$ br. wings, neb em \\  
  M31-004043.10 &   00:40:43.09	41:08:45.9  &  13 &  Fe II Em. Line(sgB[e])   & H em br wings,[Fe II] em  \\
  M31-004051.59 &   00:40:51.58	40:33:02.9  &  13,15 &  LBVc/Intermed Type  &  early A-type, see \S {3.1}, (Paper III) \\  
  M31-004052.19 &   00:40:52.18	40:31:16.5  &  13 &  Intermed Type   &  A0 I         \\
  M31-004056.49 &   00:40:56.48	41:03:08.6  &  13 &  Of/late-WN       &  H em, br.wings, He I em, He II4686, (Ofpe/WN9 \citet{Massey07})\\ 
  M31-004057.03 &   00:40:57.02	40:52:38.5  &   13 & Fe II Em. Line(sgB[e])   &  pec H$\alpha$ profile, [Fe II] em  \\
  M31-004109.26 &   00:41:09.25	40:49:05.9  &  13 &  Hot Supergiant   &  B8 I, H, neb em    \\
  M31-004130.37 &   00:41:30.36	41:05:00.8  &  13 &  Of/late-WN       &  H, neb em., He II 4686, (WNL \citet{Massey07}) \\
  M31-004220.31 &   00:42:20.30	40:51:23.1  &  13 &  Fe II Em. Line(sgB[e])   &  low S/N, H$\alpha$ br. wings, [Fe II] em \\
  M31-004253.42 &   00:42:53.41	41:27:00.4  &  13 &  Hot Supergiant   &  B5 I, H em P Cyg, br. wings \\
  M31-004259.31 &   00:42:59.30	41:06:29.0  &  13 &  forgrd dwarf     &   G:V  \\
  M31-004303.21 &   00:43:03.20	41:04:33.7  &  13 &  Hot Supergiant   &  B1 I, H em P Cyg,, neb em. \\
  PSO-J10.8180+41.6265 &  00:43:16.32	41:37:30.5 &  15 & \nodata  &  low S/N, A-type, molecular bands in red  \\
  M31-004339.28 &   00:43:39.27	41:10:19.3  &  13 &  Hot Supergiant   &  O9 I(He II 4200) star, strong H, neb em. \\
  M31-004410.90 &   00:44:10.89	41:32:03.1  &  13 &  Hot Supergiant   &  B1 I, superposed H em, neb em.\\
  PSO-J11.0457+41.5548 &  00:44:11.00	41:33:17.5 &  15 & Intermed Type  & early A, H em, He I abs, strong neb em ,[O I] neb\\
  M31-004416.28 &   00:44:16.27	41:21:06.5  &  13 &    peculiar        &  mid B + ?, molecular bands?,H em \\
  M31-004433.58 &   00:44:33.57	41:52:47.8  &  13 &  Hot Supergiant   &  B5 I , H$\alpha$ asymmetric, P Cyg, He I em \\
  M31-004434.65 &   00:44:34.64	41:25:03.5  &  13 &  Hot Supergiant   &  B2 I, H, neb em.  \\
  M31-004442.07 &   00:44:42.06 +41:27:32.2  &  13 & Hot Supergiant   &  B5 I, H em, br. wings, H$\beta$ P Cyg, neb em \\ 
  M31-004443.57 &   00:44:43.56	41:26:16.4  &  13 &  Hot Supergiant   &  B1 I, H em. P Cyg, neb em  \\
  M31-004500.90 &   00:45:00.89	41:31:00.6  &  13 &   H II             &  (WC, Massey)   \\   
  PSO-J11.2574+42.0498 &  00:45:01.83	42:02:59.0 &  15 &  Hot Supergiant & B:, H em, He I abs, molecular bands in red\\ 
  M31-004511.60 &   00:45:11.59	41:37:16.7  &  13 &   H II             &                \\
  M31-004621.08\tablenotemark{a} &   00:46:21.07 42:13:08.0 & 14  &  Warm Hypergiant &  Late B/early A, H em  P Cyg, He I, Fe II, Ca II, [Ca II] em, (Paper III)\\ 
   &               &   M33        &        &                         \\
   V-000029     &  01:31:48.11	30:32:06.8  &  13 & foreground        &  G:V \\ 
   V-001429     &  01:32:27.78	30:21:46.7  &  13 & foreground        &  G:V \\ 
   M33C-2976    &  01:32:29.0	30:28:19.5  &  13 & Fe II Em. Line    &  early type B, weak Fe II em, low S/N \\
   V-001705     &  01:32:29.05	30:34:04.1  &  13 & Intermed Type  &   A0 I   \\ 
   V-002627     &  01:32:31.91  30:35:16.7	&  14 & Intermed Type  &  A2:I, strong neb em \\
   M33C-1343    &  01:32:39.61	30:24:51.8  &  13 & Hot Supergiant &  B0 I, H, neb em \\
   V-006389     &  01:32:41.27	30:22:31.1  &  13 & Hot Supergiant &  B3 I, neb em   \\  
   V-008043     &  01:32:44.59	30:34:59.4  &  13 &   H II            &  early type B, H em P Cyg, neb em \\ 
   V-008581     &  01:32:45.51	30:39:06.6  &  13 &   H II         &   neb em \\
   M33C-13319   &  01:32:53.19	30:38:56.0  &  13 &  \nodata          &  H em only, low S/N \\
   V-013846     &  01:32:54.34  30:30:50.5  &  14 & Intermed Type  &  A2 I \\ 
   M33C-9519    &  01:32:56.32	30:35:35.3  &  13 & Of/late-WN       &   H$\alpha$ br. wings, He II 4686 (WN6 \citet{Neugent})) \\
   V-015651     &  01:32:56.44  30:35:30.8  &  14 & Hot Supergiant &  B2 I, neb em  \\
   V-017442     &  01:32:58.68	30:31:52.8  &  13 &  Hot Supergiant &  B2 I, H$\alpha$ em \\
   M33C-4444    &  01:32:59.85	30:30:43.6  &  13 & Hot Supergiant &  B I, H em, neb em  \\
   M33C-4146    &  01:33:00.17	30:30:15.2  &  13 & Of/late-WN       & H, He I, He II em, H$\alpha$ br, asymmetric wings, (WN8 \citet{Massey98})  \\
   M33-013300.86 & 01:33:00.83	30:35:04.8  &  14 & Hot Supergiant &  B1 I, H em \\
   V-021189     &  01:33:03.28  30:11:21.7  &  14 &  H II           &  neb em \\ 
   V-021331     &  01:33:03.47	30:33:23.0  &  13 &  Hot Supergiant &  B1 I, H em  \\
   M33C-3109    &  01:33:07.28	30:28:35.2  &  13 & Intermed Type  &  A5 I, H em  \\
   V-024824     &  01:33:07.57	30:42:58.9  &   13 & pec            & molecular bands?, neb em \\
   V-024835     &  01:33:07.58  30:42:42.9  &  14 & Hot Supergiant & early B type, neb em \\ 
   V-027321     &  01:33:10.40  30:38:49.3  &  14 & Hot Supergiant & B8-A0 I, H em,  H$\alpha$ br wings \\
   V-028115     &  01:33:11.23	30:45:15.2  &   13 & H II           &  neb em \\
   M33C-1141    &  01:33:11.74	30:24:23.0  &  13 &  \nodata        &  low S/N,  H em, H$\alpha$ br. wings \\
   M33C-13233   &  01:33:11.82	30:38:52.6  &  13 & Of/late-WN     &  He II 4686, asymmetric wings on H, He I, neb em (WNL, \citet{Massey98}) \\
   V-030009     &  01:33:13.14	31:04:59.2   &  13 & Foreground  &  HB, early A \\
   V-031584     &  01:33:14.78  30:45:59.2 &  14 &  Intermed Type  & early A type, H em superposed on abs, neb em \\ 
   M33-013315.21 & 01:33:15.18	30:53:18.4  &  13 &  H II          &  He I abs,He II 4686 em, neb em \\ 
   M33C-19088   &  01:33:15.28	30:45:03.4  &  14 &  WN            & early WN, strong, broad N em (WN3, \citet{Massey98})\\
   V-032629     &  01:33:15.79	30:56:44.7  &   14 & Of/late-WN     &  H$\alpha$ broad wings (WN4.5 + O \citet{Neugent})\\
   M33C-20882   &  01:33:16.34	30:46:46.6  &  13 &  H II          &  He I abs, neb em, low S/N  \\
   V-033824     &  01:33:16.89	30:23:07.2  &  13 & Hot Supergiant &  B3 I, H$\alpha$ em \\
   M33C-25255   &  01:33:17.02	30:53:29.8  &  13 &  Fe II Em. Line &  weak Fe II em  \\
   M33-013317.22 & 01:33:17.19	30:32:01.5  &  13 &  Fe II Em. Line &  B5 I, H em, Fe II, [Fe II] \\
   M33C-23380   &  01:33:24.01	30:50:30.6  &  13 &  Of/late-WN   &  He I P Cyg, He II 4686 em  (WN8, Massey) \\
   V-051296     &  01:33:30.96	30:36:52.4   &  13 &  Foreground   &  HB A-type, molecular bands? \\ 
   V-052581     &  01:33:31.77	30:22:59.0  &  13 &  Foreground   &  G: V \\
   M33C-16063   & 01:33:32.79	30:41:45.9  &  13 &   Of/late-WN   &  H br asymmetric wings, He II 4686 em, pec profiles, (WN7, \citet{Drissen}) \\
   M33C-15894   &  01:33:32.94	30:41:36.0  &  14 &  Of/late-WN   &  He II 4686 em, H, He I PCyg, very br. wings (WNL, \citet{Massey98})\\ 
   M33-013334.06 & 01:33:34.03	30:47:44.2  &  13 &  Of/late-WN  & O9 If/WN H, He I em  P Cyg, He II 4686 em, \\
   M33C-5665    &  01:33:34.36	30:32:08.3  &  13 &  Hot Supergiant &  B0 I, neb em. \\
   V-057412     &  01:33:34.58	30:40:56.1  &  13 &   Foreground   & superposed neb em \\
   M33-013335.32 & 01:33:35.29	30:39:30.9  &  13 &  Hot Supergiant &  He I P Cyg, strong neb em \\
   V-060906      &  01:33:36.51	30:20:58.1  &  13 &  Hot Supergiant &  B8 I    \\
   M33C-23421   & 01:33:36.68	30:50:35.9   &  13 &  H II           &  He I P cyg, neb em \\
   M33C-7024    & 01:33:37.32	30:33:28.9  &  13 &  Fe II Em. Line &  He I, abs, em, Fe II, [FeII] em  \\
   V-062775     & 01:33:37.55	30:28:04.6  &  13 &   Foreground     &  HB A-type  \\
   M33C-6153    & 01:33:38.96	30:32:36.2  &  13 &  Hot Supergiant &  B1 -2 I, neb em \\
   M33-013339.08 & 01:33:39.05  30:20:10.6  &  14 & Hot Supergiant &  B3 I, neb em  \\
   V-065935     & 01:33:39.22	30:43:03.5  &  13 &  Hot Supergiant &  B0 I, H em superposed on abs, neb em \\
   M33C-12405   & 01:33:39.39	30:38:10.7  &  13 &  Hot Supergiant &  O star (O8:), He II 4200, 4542, H, He I P Cyg, neb em \\
   M33-013339.42 & 01:33:39.39	30:31:24.7  & 14 &  Hot Supergiant &  B2 I, neb em \\
   V-069329     & 01:33:40.79	30:31:32.5  & 13 &   Hot Supergiant &  B5 I, neb em \\ 
   M33-013342.03 & 01:33:42.00	30:47:33.5  &  13 &   H II           &  early type B, neb em, He I em  \\
   V-072150      &  01:33:42.05	30:42:00.2  &  13 &  Hot Supergiant &  early type B, neb em \\ 
   M33C-6545    & 01:33:42.23	30:33:01.5  &  13 &  H II           &  He I em, neb em \\
   M33C-6448    & 01:33:42.75	30:32:56.2  &  14 &  Fe II Em. Line(sgB[e])  & weak Fe II, [FeII], He I em, H$\alpha$ wings \\
   M33C-13389   & 01:33:43.18	30:39:00.4  &  14 &  WC              & v broad em. (WC5 \citet{Abbott}) \\
   V-075005     & 01:33:43.31     30:35:33.8 & 14 &  Hot Supergiant &  B3 I \\
   V-077826     & 01:33:44.53	30:32:01.2  &  13 &  Hot Supergiant &  B3 I   \\
   M33C-9851    & 01:33:44.62	30:35:53.1  &  13 &  Hot Supergiant &  B1 I, neb em \\
   V-078287     & 01:33:44.72	30:44:44.4  & 13 &   Hot Supergiant &  neb em, He I em \\
   M33C-10473   & 01:33:45.22	30:36:26.5  &  14 &  Hot Supergiant &  B1 I, H em \\
   V-080679     & 01:33:45.83   30:44:44.4 &  14 &  Hot Supergiant &  B8 I, H em superposed on abs, neb em \\
   V-083744     & 01:33:47.30	30:33:06.7  &  14 &   Hot Supergiant &  B1 I, neb em \\
   M33C-17953   & 01:33:47.64	30:43:51.1  &  13 &  Of/late-WN     & H$\alpha$ very br wings,He II 4686, 5411 em, (WN6, \citet{Massey98}) \\
   V-084795     &  01:33:47.79     30:43:24.8 &   14 & Foreground     &  HB, A \\
   M33C-18822   & 01:33:47.89	30:44:48.7  &  14 &  Hot Supergiant & early type B, H em superposed on abs, neb em \\
   V-086876     & 01:33:48.86	30:21:48.5  &  13 &   Foreground     &  G: V \\
   M33C-24812   &  01:33:49.25	30:52:50.1  &  13 &  Fe II Em. Line(sgB[e]) &  Fe II, [Fell] \\
   M33C-12863   &  01:33:49.30	30:38:35.0  &  13 &  Hot Supergiant & H em, low S/N \\
   V-088927     & 01:33:49.91	30:29:28.7  &  13 &   \nodata       & H em \\
   M33C-14160   & 01:33:50.83	30:39:45.8   &  13 &  Intermed Type  &  early A, H em superposed on abs., neb em. \\
   M33-013350.92  & 01:33:50.89	30:39:36.8  &  13 &  H II           &  strong neb em \\
   M33C-7795    & 01:33:51.21	30:34:13.2   &  13 &  Intermed Type  &  early A, low S/N   \\
   M33C-15345   & 01:33:51.72	30:41:04.0  &  13 &  Hot Supergiant &  B0 I, H em \\
   M33C-17472   & 01:33:52.10	30:43:19.2  &  13 &  Hot Supergiant & early type B, H, He I PCyg, low S/N \\
   V-092983     & 01:33:52.16	30:36:36.5  &  13 &  Intermed Type  &  A0 I, H em \\
   M33C-13767   & 01:33:52.36	30:39:20.8  &  14 &  Hot Supergiant & early B, wk WN N em feature, H, He I em \\
   V-094256     & 01:33:52.92	30:44:56.9   &  13 &  Foreground     &  G: V \\  
   V-096860     & 01:33:54.61   30:33:08.1 &  14 &  Hot Supergiant &  H em , neb em \\
   M33C-5916    & 01:33:54.82	30:32:22.7  &  14 &  Of/late-WN     & He II 4686 em, He I PCyg, (WN9-10, \citet{Abbott})\\
   M33C-22178   & 01:33:55.78	30:48:31.3  &  13 &  Intermed Type  &  late B/early A, low S/N \\
   V-100400     & 01:33:56.83   30:34:29.5 &  14 &  Hot Supergiant & neb em, H em superposed on abs, He I em  \\ 
   V-100647     & 01:33:56.97   30:38:26.3 &  14 &  Intermed Type  & A0 I, H em superposed on abs \\  
   V-101408     & 01:33:57.42	30:32:27.5  &  13 &  Intermed Type  & A2 I \\
   M33C-16518   & 01:33:57.43	30:42:11.3  &  14 &  H II           & strong neb em, He I em \\
   V-102367     & 01:33:57.98   30:41:22.2   & 14 &   Foreground     &  G: V \\  
   V-103164     & 01:33:58.4	30:33:01.5  &  13 &  Hot Supergiant &  B8 I   \\
   M33C-14430   & 01:33:58.74	30:40:04.3  &  13 &  Hot Supergiant & neb em, low S/N \\
   M33C-8293    & 01:33:59.08	30:34:37.1  &  13 &  Hot Supergiant & B3 I, neb em  \\
   V-104958     & 01:33:59.37	30:23:10.9  &  13 &  Intermed Type  & B8 - A0 I  \\ 
   V-105786     & 01:33:59.85   30:33:54.8 &  14 &  Hot Supergiant & neb em, H em superposed on abs, He I em  \\   
   V-106177     & 01:34:00.07	30:46:14.9  &  13 &  Hot Supergiant   & B8 I, \\
   V-106653     & 01:34:00.56	30:37:18.4  &  13 &   H II           & late B type: low S/N, neb em   \\
   M33C-10334   & 01:34:01.13	30:36:18.3  &  13 &  Hot Supergiant & B0: I, H, He I em \\
   M33C-14422   &  01:34:01.24	30:40:03.9  &  14 &  early WN      & strong, broad N em feature \\
   M33C-9826    &  01:34:01.59	30:35:52.0  &  14 &  Hot Supergiant &  neb em \\
   M33C-11459   &  01:34:01.65	30:37:19.9  &  14 &  Hot Supergiant &  mid type B, He I em superposed on abs, neb em \\ 
   V-109457     &  01:34:02.18	30:38:50.2  &  13 &  Hot Supergiant &  B2 I, He I em, neb em \\
   V-115375     &  01:34:06.80  30:47:22.3 &  14 &    H II           &  He I em, neb em \\
   M33C-20733   & 01:34:10.58	30:46:37.7  &  13 & Hot Supergiant &  B2-3 I, H em superposed on abs. \\
   M33C-10452   &  01:34:11.53	30:36:25.3  &  14 &  Hot Supergiant &  O9, H em, P Cyg, H$\alpha$ br. wings, neb em \\
   V-123649     &  01:34:14.18	30:53:55.1  & 13 &   Intermed Type  &  red only, strong O I 7774 \\
   V-123651     &  01:34:14.18	30:33:43.2  &  13 &    \nodata      &  red only, neb em \\
   M33C-8714    & 01:34:14.87	30:34:57.3  &  14 &  Hot Supergiant &  early type B, H em superposed on abs, strong neb em \\
   M33C-11284   & 01:34:15.04	30:37:09.8  &  13 & H II/late-WN    &  neb em, He I, He II 4686 em, wk WN feature \\
   M33C-7545    &  01:34:15.7	30:34:00.6  &  14 &  Of/late-WN     & He II 4686 em, He I PCyg, (WNL \citet{Massey98}) \\
   M33C-10788   & 01:34:16.04	30:36:42.0  &  13 &  Of/late-WN     & He II 4686 em, He I PCyg, (Ofpe/WN9 \citet{Neugent}) \\
   B526SW\tablenotemark{a}  &  01:34:16.02 30:33:44.1 &  14 & Hot Supergiant &  B5 I, see \S {3.1} \\
   B526NE\tablenotemark{a}  &  01:34:16.07 30:33:44.8 &  14 & Hot Supergiant & LBV candidate:, but see \S {3.1} \\
   M33C-11332   &  01:34:16.32	30:37:12.2  &  14 &  Of/late-WN     &  He II 4686 em, H  br wings (WN7, \citet{Massey98}) \\ 
   M33C-14120   &  01:34:20.65	30:39:42.6  &  13 &  Intermed Type  &  A0 I, H em  \\
   V-130270     &  01:34:20.92  30:30:39.8 &   14 & Intermed Type  &  A8 I, H em \\ 
   M33C-20109   &  01:34:27.08	30:45:59.7  &  13 &  Fe II Em  Line & H em P Cyg, weak Fe II em, low S/N \\
   V-135855     &  01:34:29.44	30:53:12.2  &  14 &   \nodata      & neb em, low S/N \\
   V-136261     &  01:34:30.26	30:40:39.7  &  13 &  Hot Supergiant & B1-2 I    \\
   M33C-22022   &  01:34:32.02	30:48:17.0  &  13 &   Hot Supergiant & early type B, low S/N  \\ 
   M33C-21057   &  01:34:33.07	30:46:58.9  &  14 &     H II        &  He I em, strong, neb em \\
   V-139873     &  01:34:37.25	30:38:17.7   & 13 &   Hot Supergiant &  late B, strong O I 7774, H$\alpha$ br wings \\
   M33-013438.76 & 01:34:38.73	30:43:58.7  &  14 &  Hot Supergiant &  early type B, H em superposed on abs, neb em  \\
   M33C-16364   &  01:34:59.36	30:42:01.1  &  13 &  Of/late-WN     & He II 4686 em, N II em (O8Iaf, \citet{Massey95}) 
\enddata
\tablenotetext{a}{ LBT/MODS spectrum, Nov. 2014, see text. }
\end{deluxetable}

\begin{deluxetable}{llllllllllllllllll}
\rotate
\tablewidth{0 pt}
\tabletypesize{\scriptsize}
\tablenum{4-on line} 
\tablecaption{Multi-Wavelength Photometry }
\tablehead{
\colhead{Star} & 
\colhead{U}  &
\colhead{B} &
\colhead{V} &
\colhead{R} & 
\colhead{I} &
\colhead{J} &
\colhead{H} &
\colhead{K} &
\colhead{3.6$\mu$m}\tablenotemark{a} &
\colhead{4.5$\mu$m}\tablenotemark{a} &
\colhead{5.8$\mu$m}\tablenotemark{a} &
\colhead{8$\mu$m}\tablenotemark{a}  &
\colhead{3.4$\mu$m}\tablenotemark{b} &
\colhead{4.6$\mu$m}\tablenotemark{b} &
\colhead{12$\mu$m}\tablenotemark{b} &
\colhead{22$\mu$m}\tablenotemark{b} &  
\colhead{Var}
}

\startdata
   &    &   &    &  & & & & M31   & &   &  &  &  &   & &   \\ 

M31-003910.85 & 17.42 & 18.46 & 18.18 & 17.81 & 17.54 & \nodata & \nodata &\nodata & \nodata & \nodata &\nodata &\nodata & \nodata & \nodata &\nodata &\nodata \\
M31-003944.71 & 17.44 & 18.35 & 18.20 & 18.07 & 17.95 & \nodata & \nodata &\nodata & \nodata & \nodata &\nodata &\nodata & \nodata & \nodata &\nodata &\nodata \\
PSO-J10.1165+40.7082  & 18.50 & 19.54 & 19.51 & 18.94 & 17.55 & 15.62 & 14.73 & 14.32 & 13.58 & 13.19 & 12.88 & 12.03 & 13.82 & 13.56 & 10.37 & 7.92 &  var.\\ 
M31-004030.28 & 16.18 & 17.31 & 17.36 & 17.31 & 17.26 & 16.72 & 15.63 & 15.38 & \nodata & \nodata &\nodata &\nodata & \nodata & \nodata &\nodata &\nodata \\ 
M31-004032.37 & 17.49 & 18.20 & 17.76 & 17.45 & 17.09 & 16.32 & 15.69 & 15.17 & 15.18 & 14.90 & 11.97 & 10.16 &  \nodata & \nodata &\nodata &\nodata \\ 
M31-004033.80 & 16.42 & 17.36 & 17.33 & 17.23 & 17.25 & \nodata & \nodata &\nodata & \nodata & \nodata &\nodata &\nodata & 14.88 & 14.92 & 10.77 & 8.65\\
M31-004043.10 & 17.70 & 18.77 & 18.62 & 17.56 & 17.72 & 17.44 & 15.82 & 15.30 & 14.90 & 14.60 & 13.75 & 13.38 & 13.56 & 12.58 & 10.50 & 8.55\\
M31-004051.59 & 16.44 & 17.21 & 16.99 & 16.77 & 16.58 & 16.38 & 15.59 & 15.75 & \nodata & \nodata &\nodata &\nodata & 14.43 & 12.30 & 9.92 & 8.21\\
M31-004052.19 & 17.28 & 18.02 & 17.69 & 17.48 & 17.25 & \nodata & \nodata &\nodata & \nodata & \nodata &\nodata &\nodata & \nodata & \nodata &\nodata &\nodata \\
M31-004056.49 & 17.01 & 18.00 & 18.09 & 18.00 & 17.89 & \nodata & \nodata &\nodata & \nodata & \nodata &\nodata &\nodata & \nodata & \nodata &\nodata &\nodata \\
M31-004057.03 & 18.89 & 19.21 & 18.81 & 18.46 & 18.13 & \nodata & \nodata &\nodata & \nodata & \nodata &\nodata &\nodata & \nodata & \nodata &\nodata &\nodata \\
M31-004109.26 & 17.13 & 17.67 & 17.36 & 17.11 & 16.85 & 16.41 & 15.96 & 15.09 & 13.23 & 13.14 & 12.55 & 11.92 & 13.28 & 13.17 & 10.02 & 7.37 \\
M31-004130.37 & 17.53 & 18.55 & 18.50 & 18.12 & 18.17 & \nodata & \nodata &\nodata & 15.45 & 15.09 & 12.20 & 10.49 & \nodata & \nodata &\nodata &\nodata  \\
M31-004220.31 & 18.80 & 19.15 & 18.90 & 18.56 & 18.34 & \nodata & \nodata &\nodata & 13.97 & 13.28 & 12.60 & 11.65 & 14.18 & 13.30 & 10.52 & 8.81 \\ 
M31-004253.42 & 16.39 & 17.28 & 17.27 & 17.22 & 17.20 & \nodata & \nodata & \nodata & \nodata & \nodata &\nodata &\nodata & \nodata & \nodata & \nodata &\nodata \\
M31-004259.31 & 17.65 & 17.78 & 17.07 & 16.71 & 16.21 & 16.10 & 15.51 & 15.56 & \nodata & \nodata &\nodata &\nodata & \nodata & \nodata & \nodata &\nodata \\
M31-004303.21 & 17.51 & 18.21 & 17.84 & 17.57 & 17.29 & \nodata & \nodata &\nodata & 15.13 & 14.95 & 14.10 & 13.01 & 14.94 & 14.40 & 10.16 & 7.59 \\
PSO-J10.8180+41.6265  &  19.18 & 19.66 & 19.49 & 17.82 & 16.52 & 14.67 & 13.90 & 13.46 & 12.82 & 12.93 & 12.65 & 12.19 & 13.18 & 13.40 & 10.59 & 8.19&  var. \\ 
M31-004339.28 & 17.98 & 18.90 & 18.86 & 18.69 & 18.73 & \nodata & \nodata &\nodata & \nodata & \nodata &\nodata &\nodata & 14.46 & 13.92 & 9.48 & 7.18 \\
M31-004410.90 & 16.50 & 17.42 & 17.48 & 17.47 & 17.46 & \nodata & \nodata &\nodata & \nodata & \nodata & \nodata & \nodata & \nodata & \nodata & \nodata & \nodata \\
PSO-J11.0457+41.5548  & 16.55 & 17.34 & 17.30 & 17.11 & 16.96 & 16.48 & 15.09 & 14.88 & \nodata & \nodata & \nodata & \nodata & 13.92 & 13.52 & 9.20 & 7.08 &  var. \\
M31-004416.28 & 18.27 & 18.58 & 18.35 & 17.94 & 16.50 & 14.94 & 14.14 & 13.60 & 13.04 & 12.93 & 11.86 & 10.37 & 12.99 & 12.80 & 8.86 & 6.83 \\ 
M31-004433.58 & 17.89 & 18.45 & 18.16 & 17.80 & 17.55 & 16.83 & 16.80 & 15.56 & \nodata & \nodata &\nodata &\nodata & 13.35 & 12.71 & 7.92 & 5.56 \\
M31-004434.65 & 15.71 & 16.61 & 16.66 & 16.63 & 16.66 & 16.53 & 16.42 & 15.13 & \nodata & \nodata &\nodata &\nodata & \nodata & \nodata &\nodata &\nodata &  var.\\
M31-004442.07 & 17.64 & 18.41 & 17.98 & 17.65 & 17.36 & \nodata & \nodata & \nodata & \nodata & \nodata &\nodata &\nodata & 14.98 & 14.02 & 9.09 & 6.93 \\ 
M31-004443.57 & 16.71 & 17.64 & 17.71 & 17.67 & 17.74 & \nodata & \nodata &\nodata & \nodata & \nodata &\nodata &\nodata & \nodata & \nodata &\nodata &\nodata \\
M31-004500.90 & 18.35 & 19.19 & 19.05 & 18.60 & 18.63 & \nodata & \nodata & \nodata & \nodata & \nodata &\nodata &\nodata & 13.41 & 12.85 & 8.05 & 5.81 \\
PSO-J11.2574+42.0498 & 17.86 & 18.84 & 18.50 & 17.84 & 16.76 & 15.24 & 14.50 & 14.15 & 13.52 & 13.61 & 13.29 & 12.75 & 13.80 & 13.82 & 10.96 & 9.27 & var. \\
M31-004511.60 & 18.06 & 19.06 & 19.11 & 18.93 & 19.22 & 16.75 & 15.71 & 14.97 & 14.39 & 14.08 & 11.21 & 9.65 & 12.49 & 12.05 & 7.56 & 5.50\\ 
M31-004621.08 & 17.88 & 18.45 & 18.16 & 17.72 & 17.39 & 16.81 & 16.43 & 15.15 & 13.42 & 12.46 & 12.21 & 11.09 & 13.74 & 12.65 & 10.12 & 8.34\\

   &               &              &                               \\ 
      &    &   &    &  & & & & M33   & &   &  &  &  &   & &   \\
M33C-2976   & 18.27  & 19.04 & 19.00  & 18.87 & 18.76  & \nodata & \nodata & \nodata & 16.59   & 16.09    & \nodata & \nodata & \nodata & \nodata & \nodata & \nodata \\
V-001705    & 17.41  & 18.07 & 18.02 & 17.97 & 17.88  & 15.95   & 15.03   & 14.92   & 14.76   & 14.97   & \nodata & \nodata & 14.54 & 14.82 & 12.31 & 8.86 \\
V-002627    & 18.47  & 18.11 & 17.39 & 17.00 & 16.55  & 15.82   & 15.51   & 15.33   & 15.28    & 15.07 & \nodata & \nodata & 14.51 & 14.14 & 9.83 & 7.48 \\
M33C-1343   & 17.14  & 18.21 & 18.32 & 18.38 & 18.47  & \nodata & \nodata & \nodata & \nodata & \nodata & \nodata & \nodata & \nodata & \nodata & \nodata & \nodata \\
V-006389    & 17.01  & 17.97 & 18.01 & 17.97 & 17.98  & \nodata & \nodata & \nodata & \nodata & \nodata & \nodata & \nodata & \nodata & \nodata & \nodata & \nodata \\
V-008043    & 16.84  & 18.14 & 18.15 & 18.04 & 18.21  & \nodata & \nodata & \nodata & 15.63 & \nodata & \nodata & 10.97 & 15.07 & 14.44 & 9.69 & 7.03 \\
V-008581    & 18.21  & 19.22 & 19.31 & 18.83 & 18.27  & \nodata & \nodata & \nodata & 16.24    & 16.27 & \nodata & \nodata & \nodata & \nodata & \nodata & \nodata \\
M33C-14239  & 16.13  & 17.32 & 17.25 & 17.01 & 16.89  & 15.68   & 15.28   & 15.06   & 14.85 & 14.76   & \nodata & \nodata & 14.88 & 14.65 & 13.10 & 9.47 \\
M33C-13319  & 19.11  & 19.74 & 19.36 & 19.08 & 18.76  & \nodata & \nodata & \nodata & \nodata & \nodata & \nodata & \nodata & \nodata & \nodata & \nodata & \nodata \\
V-013846    & 17.61  & 17.95 & 17.65 & 17.47 & 17.23  & 16.71   & 16.17   & 16.30   & 16.57   & 16.67   & \nodata & \nodata & 16.81 & 16.21 & 11.60 & 9.14 \\
M33C-9519   & 17.90  & 18.74 & 18.93 & 18.86 & 18.89  & \nodata & \nodata & \nodata & \nodata & \nodata & \nodata & \nodata & \nodata & \nodata & \nodata & \nodata \\
V-015651    & 17.26  & 18.33 & 18.41 & 18.42 & 18.39  & \nodata & \nodata & \nodata & \nodata & \nodata & \nodata & \nodata & \nodata & \nodata & \nodata & \nodata \\
V-017442    & 17.02  & 17.95 & 17.82 & 17.72 & 17.58  & \nodata & \nodata & \nodata & 17.35   & 17.35 & \nodata & \nodata & 16.06 & 16.01 & 9.79 & 7.86 \\
M33C-4444   & 17.68  & 18.75 & 18.84 & 18.86 & 18.91  & \nodata & \nodata & \nodata & \nodata & \nodata & \nodata & \nodata & \nodata & \nodata & \nodata & \nodata \\
M33C-4146   & 17.66  & 18.70 & 18.79 & 18.71 & 18.67  & \nodata & \nodata & \nodata & \nodata & \nodata & \nodata & \nodata & \nodata & \nodata & \nodata & \nodata \\
M33-013300.86 & 16.11 & 17.20 & 17.32 & 17.36 & 17.38 & 15.85 & 15.14 & 14.77 & 14.80 & 14.94 & \nodata & \nodata & 14.56 & 14.23 & 10.14 & 7.71 \\
V-021189    & 17.26  & 18.22 & 18.36 & 18.39 & 18.41  & \nodata & \nodata & \nodata & 15.78 & \nodata & 11.21 & \nodata & 15.01 & 14.57 & 10.15 & 7.10 \\
V-021331    & 17.56  & 18.38 & 18.47 & 18.45 & 18.43  & \nodata & \nodata & \nodata & \nodata & \nodata & \nodata & \nodata & \nodata & \nodata & \nodata & \nodata \\
M33C-3109   & 19.22  & 19.19 & 18.78 & 18.55 & 18.31  & \nodata & \nodata & \nodata & 17.45   & 17.38   & \nodata & \nodata & \nodata & \nodata & \nodata & \nodata \\
V-024824    & 17.45  & 19.81 & 18.36 & 17.20 & 16.00  & 14.32   & 13.52   & 13.20   & 13.07   & 12.99    & \nodata & 11.97 & 12.76 & 12.60 & 10.49 & 8.90 \\
V-024835    & 17.03  & 18.16 & 18.32 & 18.39 & 18.47  & \nodata & \nodata & \nodata & \nodata & \nodata & \nodata & \nodata & \nodata & \nodata & \nodata & \nodata \\
V-027321    & 16.86  & 17.76 & 17.68 & 17.56 & 17.42  & 16.74 & 15.92 & 16.06 & 14.62   & 13.56   & \nodata & 11.26 & 14.70 & 13.39 & 9.36 & 6.84 \\
V-028115    & 15.32  & 16.36 & 16.24 & 16.17 & 16.15  & 15.54   & 15.27   & 14.91   & 13.44 & 13.06   & \nodata & \nodata & 12.82 & 11.96 & 7.20 & 4.11 \\
M33C-1141   & 18.34  & 19.42 & 19.39 & 19.27 & 19.37  & \nodata & \nodata & \nodata & 17.35   & 17.27   & \nodata & \nodata & \nodata & \nodata & \nodata & \nodata \\
M33C-13233     & 17.01  & 17.96 & 18.14 & 18.06 & 18.05  & \nodata & \nodata & \nodata & 14.30 & 14.31 & \nodata & \nodata & 13.92 & 12.88 & 8.09 & 5.07 \\
V-031584    & 18.60  & 18.71 & 18.24 & 17.97 & 17.68  & \nodata & \nodata & \nodata & \nodata & \nodata & \nodata & \nodata & \nodata & \nodata & \nodata & \nodata \\
M33-013315.21 & 17.60 & 18.54 & 18.53 & 18.45 & 18.45 & 14.95 & 14.24 & 13.83 & 13.19 & 13.08 & \nodata & 11.40 & 13.06 & 12.82 & 9.70 & 7.22 \\
M33C-19088     & 17.53  & 18.40 & 18.53 & 18.46 & 18.67 & \nodata & \nodata & \nodata & \nodata & \nodata & \nodata & \nodata & \nodata & \nodata & \nodata & \nodata \\
V-032629    & 15.62  & 16.69 & 16.79 & 16.79 & 16.84  & 16.61   & 15.89    & 15.65   & 15.75   & 14.92   & \nodata & \nodata &  \nodata & \nodata &\nodata &\nodata \\
M33C-20882     & 17.48  & 18.58 & 18.66 & 18.69  & 18.76 & \nodata & \nodata & \nodata & \nodata & \nodata & \nodata & \nodata & \nodata & \nodata & \nodata & \nodata \\
V-033824    & 17.75  & 18.45 & 18.43 & 18.38 & 18.36  & 15.94   & 15.61   & 15.20   & \nodata & \nodata & \nodata & \nodata & 14.63 & 14.54 & 12.00 & 9.31 \\
M33C-25255     & 17.86  & 18.96 & 18.92 & 18.67 & 18.59  & \nodata & \nodata & \nodata & 17.57   & 17.53   & \nodata & \nodata & \nodata & \nodata & \nodata & \nodata \\
M33-013317.22 & 18.56 & 19.39 & 18.75 & 18.02 & 17.23  & 15.95 & 15.09 & 14.88 & 14.86 & 15.22   & \nodata & \nodata & \nodata & \nodata & \nodata & \nodata \\
M33C-23380     & 17.89  & 18.93 & 19.07 & 18.97 & 18.98  & \nodata & \nodata & \nodata & \nodata & \nodata & \nodata & \nodata & \nodata & \nodata & \nodata & \nodata \\
M33C-16063     & 17.04  & 17.99 & 18.20  & 18.15 & 18.17  & \nodata & \nodata & \nodata & 15.08 & 16.01 & \nodata & \nodata & \nodata & \nodata &\nodata &\nodata \\
M33C-15894     & 16.96  & 17.94 & 18.14 & 18.06 & 18.09   & \nodata & \nodata & \nodata & 16.22 & 17.53 & \nodata & \nodata & \nodata & \nodata & \nodata & \nodata \\
M33-013334.06 & 16.43 & 17.47 & 17.46 & 17.41 & 17.38 & \nodata & \nodata & \nodata & 16.80 & 16.93 & \nodata & \nodata & \nodata & \nodata &\nodata &\nodata \\
M33C-5665      & 17.26  & 18.32 & 18.39  & 18.32 & 18.41  & 16.35   & 15.90    & 15.42   & \nodata & \nodata & \nodata & \nodata & 13.34 & 12.67 & 7.53 & 5.29 \\
M33-013335.32 & 18.85 & 19.47 & 19.38 & 19.21 & 19.13 & \nodata & \nodata & \nodata & 12.84 & 12.75 & \nodata & 7.89 & 12.94 & 12.21 & 6.91 & 3.99 \\
V-060906    & 17.92  & 18.40 & 18.40 & 18.32 & 18.36  & \nodata & \nodata & \nodata & 17.00   & 17.03 & \nodata & \nodata & 16.42 & 15.58 & 11.32 & 9.55 \\ 
M33C-23421     & 17.57  & 18.51 & 18.79 & 18.61 & 18.73  & \nodata & \nodata & \nodata & 15.77   & 15.51   & \nodata & 11.46 & 15.06 & 14.42 & 9.73 & 6.84 \\ 
M33C-7024      & 17.51  & 18.45 & 18.51 & 18.46 & 18.39  & \nodata & \nodata & \nodata & \nodata & \nodata & \nodata & \nodata & \nodata & \nodata & \nodata & \nodata \\
M33C-6153      & 17.45  & 18.49 & 18.56 & 18.54 & 18.52  & \nodata & \nodata & \nodata & 14.54 & \nodata & \nodata & 10.13 & 13.17 & 13.87 & 9.13 & 6.17 \\
M33-013339.08 & 16.25 & 17.13 & 17.21 & 17.21 & 17.24 & 15.93 & 15.58 & 15.24 & 15.11 & 15.06 & \nodata & \nodata & 15.03 & 14.84 & 11.49 & 9.31 & \\
V-065935    & 17.29  & 18.33 & 18.45 & 18.52 & 18.59  & \nodata & \nodata & \nodata & \nodata & \nodata & \nodata & \nodata & \nodata & \nodata & \nodata & \nodata \\
M33C-12405     & 17.41  & 18.56 & 18.73 & 18.66 & 18.46  & \nodata & \nodata & \nodata & 14.20 & 14.80 & \nodata & 9.71 & 13.81 & 13.50 & 8.88 & 7.43\\
M33-013339.42 & 16.19 & 16.92 & 17.32 & 17.04 & 17.12 & 16.07 & 14.95 & 14.53 & \nodata & \nodata & \nodata & \nodata & \nodata & \nodata & \nodata & \nodata &  var.\\
V-069329    & 17.61  & 18.41 & 18.40  & 18.34 & 18.24  & \nodata & \nodata & \nodata & \nodata & \nodata & \nodata & \nodata & \nodata & \nodata & \nodata & \nodata \\
M33-013342.03 & 18.32 & 19.34 & 19.37 & 19.23 & 19.29 & \nodata & \nodata & \nodata & \nodata & \nodata & \nodata & \nodata & 14.32 & 13.81 & 8.92 & 6.79\\
V-072150    & 17.61  & 18.36 & 18.38 & 18.37 & 18.31  & \nodata & \nodata & \nodata & \nodata & \nodata & \nodata & \nodata & \nodata & \nodata & \nodata & \nodata \\
M33C-6545      & 17.56  & 18.71 & 18.86 & 18.88 & 18.91  & 15.12 & 14.42 & 14.05 & 14.01 & 14.30 & \nodata & 9.87 & 14.50 & 13.71 & 9.23 & 6.09 \\
M33C-6448      & 17.02  & 18.07 & 18.09 & 17.95 & 17.86  & \nodata & \nodata & \nodata & \nodata & \nodata & \nodata & \nodata & 13.54 & 13.34 & 10.80 & 6.56\\
M33C-13389 & 17.79 & 18.69 & 18.63 & 18.52 & 18.32 & \nodata & \nodata & \nodata & 14.86 & 14.78 & \nodata & \nodata & \nodata & \nodata & \nodata & \nodata \\
V-075005    & 16.74  & 17.36 & 18.02 & 17.24 & 16.43  & 15.43   & 14.64   & 14.33   & 13.95   & 14.04   & \nodata & \nodata & 13.90 & 13.85 & 9.95 & 7.36 \\ 
V-077826    & 17.27  & 18.16 & 18.21 & 18.12 & 18.13  & \nodata & \nodata & \nodata & 14.33 & 14.85  & \nodata & 9.45 & 13.79 & 13.49 & 8.66 & 6.80 \\ 
M33C-9851      & 17.95  & 18.93 & 18.90  & 18.84 & 18.75  & \nodata & \nodata & \nodata & 14.87 & 15.17 & \nodata & 10.04 & \nodata & \nodata & \nodata & \nodata &  var.\\
V-078287    & 18.72  & 19.07 & 18.48 & 18.07 & 17.59   & 16.61   & 16.25   & 15.20   & 15.07   & 15.11 & \nodata & \nodata & \nodata & \nodata & \nodata & \nodata \\
M33C-10473     & 16.61  & 17.67 & 17.82 & 17.85 & 17.84  & \nodata & \nodata & \nodata & \nodata & \nodata  & \nodata & \nodata &  \nodata & \nodata &\nodata &\nodata & var. \\
V-080679    & 18.09   & 18.27 & 18.10 & 17.99 & 17.81  & 15.37   & 14.48   & 14.22   & 13.83   & 14.12   & \nodata & \nodata & 13.09 & 13.12 & 9.64 & 6.58 \\
V-083744    & 16.83  & 17.82 & 17.87 & 17.85 & 17.81  & 16.58   & 15.82   & 14.83   & \nodata & \nodata & \nodata & \nodata & \nodata & \nodata & \nodata & \nodata &  var.\\
M33C-17953     & 17.49   & 18.47 & 18.55 & 18.51 & 18.42  & \nodata & \nodata & \nodata & 14.47 & 14.32 & \nodata & \nodata & 14.86 & 14.21 & 10.65 & 7.90 \\
V-084795    & 18.25  & 18.67 & 18.44 & 18.14 & 17.58  & 16.50 & 15.64   & 15.39   & 15.39   & 16.11 & \nodata & \nodata & 15.18 & 14.82 & 12.45 & 9.18 \\
M33C-18822     & 18.19  & 19.00 & 18.94 & 18.86 & 18.72  & 16.92   & 15.87   & 15.53   & 14.84  & 15.24   & \nodata & \nodata & 14.59 & 13.83 & 8.6 & 6.61 \\
M33C-24812     & 18.16  & 19.20  & 19.04 & 18.72 & 18.63  & \nodata & \nodata & \nodata & 14.90   & 14.34   & \nodata & 12.92 & 15.07 & 14.57 & 10.59 & 8.77 \\
M33C-12863     & 18.68  & 19.43 & 19.17 & 18.98 & 18.69   & \nodata & \nodata & \nodata & 14.53 & 14.63 & \nodata & \nodata &  \nodata & \nodata &\nodata &\nodata \\
V-088927    & 18.24  & 19.09 & 18.10 & 17.17 & 16.07  & 14.94   & 14.15   & 13.78   & 13.01 & 12.68   & \nodata & 10.54 & 12.95 & 12.59 & 9.16 & 7.73 \\
M33C-14160     & 18.73  & 18.83 & 18.62 & 18.48 & 18.30   & 16.06   & 13.84   & 13.63   & \nodata & \nodata & \nodata & \nodata & \nodata & \nodata & \nodata & \nodata \\
M33-013350.92 & 15.35 & 15.14 & 14.17 & 13.69 & 13.35 & 12.06 & 11.44 & 11.22 & 10.98 & 11.08 & \nodata & 11.40 & 10.61 & 10.63 & 9.05 & 6.56 \\
M33C-7795      & 19.04  & 19.08 & 18.81 & 18.58 & 18.38  & \nodata & \nodata & \nodata & \nodata & \nodata & \nodata & \nodata & \nodata & \nodata & \nodata & \nodata \\
M33C-15345     & 17.42  & 18.44 & 18.52 & 18.51 & 18.52  & \nodata & \nodata & \nodata & 15.41 & 15.64 & \nodata & \nodata & \nodata & \nodata & \nodata & \nodata \\
M33C-17472     & 18.19  & 18.95 & 18.72 & 18.54 & 18.34  & \nodata & \nodata & \nodata & 14.42 & 14.86 & \nodata & \nodata & 14.36 & 13.64 & 11.69 & 8.49 \\
V-092983    & 17.49  & 18.04 & 17.85 & 17.69  & 17.45  & \nodata & \nodata & \nodata & \nodata & 16.67 & \nodata & 11.20 & \nodata & \nodata & \nodata & \nodata & var. \\
M33C-13767     & 15.71  & 16.77 & 16.87 & 16.84 & 16.84  & 16.64   & 15.90  & 15.27   & \nodata & \nodata & \nodata & \nodata & \nodata & \nodata & \nodata & \nodata & var.? \\
V-096860    & 16.77  & 17.65 & 17.62 & 17.55 & 17.41  & \nodata & \nodata & \nodata & \nodata & \nodata & \nodata & \nodata & \nodata & \nodata & \nodata & \nodata &  var. \\
M33C-5916      & 17.23  & 18.34 & 18.34 & 18.21 & 18.13  & \nodata & \nodata & \nodata & 15.85   & 15.85   & \nodata & \nodata & \nodata & \nodata & \nodata & \nodata \\
M33C-22178     & 19.00  & 19.36 & 19.17 & 19.00 & 18.75  & \nodata & \nodata & \nodata & 16.79   & 17.27   & \nodata & \nodata & \nodata & \nodata & \nodata & \nodata \\
V-100400    & 18.34  & 18.97 & 19.18 & 19.23 & 18.90  & \nodata & \nodata & \nodata & \nodata & \nodata & \nodata & \nodata & \nodata & \nodata & \nodata & \nodata \\
V-100647    & 17.84  & 18.41 & 18.39 & 18.35 & 18.24  & \nodata & \nodata & \nodata & 16.69 & 17.43 & \nodata & \nodata & \nodata & \nodata & \nodata & \nodata \\
V-101408    & 18.10  & 18.30 & 18.21 & 18.14 & 18.01  & \nodata & \nodata & \nodata & \nodata & \nodata & \nodata & \nodata & \nodata & \nodata & \nodata & \nodata \\
M33C-16518     & 17.70  & 18.73 & 18.84 & 18.82 & 18.92  & \nodata & \nodata & \nodata & \nodata & \nodata & \nodata & \nodata & 13.96 & 13.40 & 8.66 & 6.34 \\
V-103164    & 18.37  & 18.47 & 18.44 & 18.42 & 18.28  & 15.67   & 15.06   & 14.63   & 14.51   & 14.85   & \nodata & \nodata & \nodata & \nodata & \nodata & \nodata & var.\\ 
M33C-14430     & 18.15  & 18.98 & 18.94 & 18.85 & 18.69  & \nodata & \nodata & \nodata & \nodata & \nodata & \nodata & \nodata & 14.06 & 13.99 & 10.68 & 9.04 \\
M33C-8293      & 16.76  & 17.75 & 17.85 & 17.87 & 17.88  & 15.92   & 15.24   & 15.14   & 14.95   & 15.50 & \nodata & \nodata & \nodata & \nodata & \nodata & \nodata \\
V-104958    & 16.48  & 17.28 & 17.18 & 17.06 & 16.99  & 16.70  & 16.01   & 15.61   & 16.31   & 16.17   & \nodata & \nodata & 15.96 & 15.71 & 12.41 & 9.42 \\
V-105786    & 16.98  & 17.98 & 18.19 & 18.26 & 18.29  & 14.38   & 13.60   & 13.19   & 12.89    & 13.02   & \nodata & 11.64 & 12.60 & 12.57 & 9.11 & 6.54 \\
V-106177    & 17.49  & 17.93 & 17.72 & 17.57 & 16.94  & 15.50   & 14.72   & 14.42   & 14.21    & 14.49   & \nodata & \nodata & 14.10 & 14.21 & 12.04 & 8.45 \\
V-106653    & 19.80  & 20.30 & 20.07 & 19.88 & 19.46  & \nodata & \nodata & \nodata & 15.60 & 16.51 & \nodata & \nodata & \nodata & \nodata & \nodata & \nodata \\
M33C-10334     & 17.47  & 18.33 & 18.23 & 18.09 & 17.94  & 16.71   & 16.02   & 15.11   & \nodata & \nodata & \nodata & \nodata & \nodata & \nodata & \nodata & \nodata \\
M33C-14422     & 17.50  & 18.48 & 18.62 & 18.58 & 18.30  & 16.87   & 16.30   & 15.45   & 14.70   & 15.07   & \nodata & 10.75 & \nodata & \nodata &\nodata &\nodata \\
M33C-9826      & 17.57  & 18.63 & 18.71 & 18.69 & 19.67  & \nodata & \nodata & \nodata & 15.14 & 14.95 & \nodata & \nodata & \nodata & \nodata &\nodata &\nodata  &   var.\\
M33C-11459     & 17.17  & 18.28 & 18.50 & 18.55 & 18.56  & \nodata & \nodata & \nodata & \nodata & 15.92 & \nodata & 11.22 & \nodata & \nodata & \nodata & \nodata \\
V-109457    & 16.99  & 18.04 & 18.13 & 18.18 & 18.19  & \nodata & \nodata & \nodata & \nodata & 16.31 & \nodata & 11.32 & \nodata & \nodata & \nodata & \nodata \\
V-115375    & 17.40  & 17.93 & 17.89 & 17.50 & 17.92  & \nodata & \nodata & \nodata & 13.23   & 12.93   & \nodata & 8.79 & 13.09 & 12.35 & 7.36 & 4.39 \\
M33C-20733     & 17.65  & 18.72 & 18.86 & 18.81 & 18.68  & \nodata & \nodata & \nodata & \nodata & \nodata & \nodata & \nodata & \nodata & \nodata & \nodata & \nodata \\
M33C-10452     & 17.56  & 18.49 & 18.36 & 18.23 & 17.99  & \nodata & \nodata & \nodata & \nodata & \nodata & \nodata & \nodata & \nodata & \nodata & \nodata & \nodata \\
V-123649    & 18.70 & 18.59 & 18.47 & 18.35 & 18.19  & \nodata & \nodata & \nodata & 15.16 & 15.45 & \nodata & \nodata & 14.60 & 14.27 & 11.27 & 9.55 \\
V-123651    & 17.26  & 18.31 & 18.39 & 18.38 & 18.37  & \nodata & \nodata & \nodata & \nodata & \nodata & \nodata & \nodata & \nodata & \nodata & \nodata & \nodata \\
M33C-8714      & 18.42  & 19.29 & 19.16 & 19.03 & 18.86  & 14.87   & 14.09   & 13.68   & 13.38   & 13.26   & \nodata & \nodata & 13.08 & 12.83 & 9.25 & 7.31 \\
M33C-11284     & 17.44  & 18.54 & 18.74 & 18.82 & 18.79  & \nodata & \nodata & \nodata & 13.84 & \nodata & \nodata & 9.63 & 13.61 & 12.88 & 7.88 & 4.49 \\
M33C-7545      & 17.51  & 18.52 & 18.67 & 18.54 & 18.47  & \nodata & \nodata & \nodata & \nodata & \nodata & \nodata & \nodata & \nodata & \nodata & \nodata & \nodata \\
M33C-10788     & 16.80  & 17.89 & 17.95 & 17.86 & 17.83  & 16.98   & 16.43   & 15.60   & 15.43   & 15.72   & \nodata & \nodata & 15.19 & 14.97 & 10.30 & 7.65 &   var.\\
B526SW        & 16.58 & 17.33 & 17.26 & \nodata & 17.09 & 16.42 & 16.44 & 15.97 & 14.57 & 15.10 & \nodata & \nodata & 13.82 & 13.84 & 10.21 & 7.18 \\
B526NE        & 16.59 & 17.33 & 17.26 & \nodata & 16.92 & 16.42 & 16.44 & 15.97 & 14.57 & 15.10 & \nodata & \nodata & 13.82 & 13.84 & 10.21 & 7.18 \\
M33C-11332     & 16.90  & 17.89 & 18.13 & 18.07 & 18.02  & \nodata & \nodata & \nodata & \nodata & \nodata & \nodata & \nodata & \nodata & \nodata & \nodata & \nodata &  var.\\
M33C-14120     & 17.54  & 18.09 & 18.00 & 17.92 & 17.76  & \nodata & \nodata & \nodata & \nodata & \nodata & \nodata & \nodata & \nodata & \nodata & \nodata & \nodata \\
V-130270    & 16.47  & 16.68 & 16.49 & 16.36 & 16.20  & 15.99   & 15.89   & 15.85   & 15.66  & 15.62   & \nodata & 13.33 & 15.41 & 14.95 & 11.29 & 8.43 \\
M33C-20109     & 17.88  & 18.56 & 18.46 & 18.33 & 18.22  & 16.38   & 15.72   & 15.46   & \nodata & \nodata & \nodata & \nodata & \nodata & \nodata & \nodata & \nodata \\
V-135855    & \nodata & 22.36 & 20.57 & 19.63 & 18.67  & \nodata & \nodata & \nodata & 16.17   & 16.41   & \nodata & \nodata & 15.86 & 15.89 & 11.47 & 8.95 \\
V-136261    & 16.76  & 17.80 & 17.88 & 17.91 & 17.94  & 15.34 & 14.04 & 13.78 & 13.53 & 13.48   & \nodata & 11.86 & 13.51 & 13.28 & 10.07 & 7.72 \\
M33C-22022     & 17.62  & 18.64 & 18.67 & 18.66 & 18.71  & \nodata & \nodata & \nodata & \nodata & \nodata & \nodata & \nodata & 15.20 & 14.49 & 9.53 & 6.86\\
M33C-21057     & 15.67  & 16.66 & 16.64 & 16.58 & 16.57  & 15.61   & 15.48   & 15.26   & \nodata & \nodata & \nodata & \nodata & \nodata & \nodata & \nodata & \nodata \\
V-139873    & 17.06  & 17.74 & 17.39 & 17.17 & 16.85  & 16.61   & 15.94   & 16.06   & 16.04   & 16.01   & \nodata & \nodata & 15.78 & 15.00 & 10.01 & 7.93 \\
M33-013438.76 & 16.38 & 17.52 & 17.66 & 17.74 & 17.87 & 16.09 & 15.46 & 15.07 & 14.38 & 14.48 & \nodata & \nodata & 15.16 & 13.71 & 8.94 & 6.27 \\
M33C-16364     & 17.35  & 18.11 & 18.25 & 18.29 & 18.41  & \nodata & \nodata & \nodata & \nodata & \nodata & \nodata & \nodata & \nodata & \nodata & \nodata & \nodata

\enddata
\tablenotetext{a}{{\it Spitzer}/IRAC}
\tablenotetext{b}{WISE}
\end{deluxetable}


\begin{thebibliography}{} 
\bibitem[Abbotot et al.(2004)]{Abbott}Abbott, J. B., Crowther, P. A., Drissen, L., Dessart, L., Martin, P., \& Boivin, G. 2004, \mnras, 350, 552 
\bibitem[Adams et al.(2016)]{Adams}Adams, S. M., Kochanek, C. S., Gerke, J. R., \& Stanek, K. Z.2016, arXiv:1610.02402
\bibitem[Aret et al.(2012)]{Aret12}Aret, A., Kraus, M., Muratore, M. F. \& Fernandes, M. B. 2012, \mnras, 423,284 
\bibitem[Aret et al.(2016)]{Aret2016}Aret, A., Kraus, M., \& Slecha, M. 2016, \mnras, 456, 1424
\bibitem[Burggraf et al.(2015)]{Burg15}Burggraf, B. et al. 2015, \aap, 581, 12  

\bibitem[Burggraf(2014)]{BurgPhD}Burggraf, B. 2014, Ph.D Dissertation  

\bibitem[Clark et al.(2012)]{Clark12}Clark, J. S., Castro, N., Garcia, M., et al. 2012, \aap, 541, A146

\bibitem[Cutri et al.(2003)]{Cutri}Cutri, R. M., Skrutskie, M. F., Van Dyk, S. et al. 2003, The IRSA 2MASS All-Sky Point Source Catalog, NASA/IPAC Infrared Science Archive 

\bibitem[Drissen et al.(2008)]{Drissen}Drissen, L., Crowther, P. A., Ubeda, L., \& Martin, P. 2008, \mnras, 389, 1033

\bibitem[Febricant et al.(2005)]{Fab}Fabricant, D., Fata, R., Roll, J., et al. 2005, \pasp, 117, 1411  

\bibitem[Gordon et al.(2016)]{Gordon}Gordon, M. S., Humphreys, R. M. \& Jones, T. J. 2016, \apj, 825, 50 (Paper III) 

\bibitem[Hartman et al.(2006)]{Hartman}Hartman, J. D., Bersier, D., Stanek, K. Z., et al. 2006, \mnras, 371, 1405 

\bibitem[Humphreys(1975)]{RMH75}Humphreys, R. M. 1975, \apj, 200, 426

\bibitem[Humphreys \& Sandage(1980)]{HS80}Humphreys, R. M \& Sandage, A. 1980, \apjs, 44, 319

\bibitem[Humphreys \& Davidson(1994)]{HD94}Humphreys. R. M. and Davidson, K. 1994, \pasp, 106, 1025 

\bibitem[Humphreys et al.(2013)]{RMH13}Humphreys. R. M., Davidson, K, Grammer, S., Kneeland, N., Martin, J. C., Weis, K. \& Burggraf, B. 2013, \apj, 773, 46 (Paper I)  

\bibitem[Humphreys et al.(2014a)]{RMH2014a}Humphreys, R. M., Davidson, K., Gordon, M. Weis, K. Burggraf, B., Bomans, D.~J. \& Martin, J.~C. 2014, \apjl, 782L, 21
\bibitem[Humphreys et al.(2014b)]{RMH2014b}Humphreys, R. M.,Weis, K.,Davidson, K., Bomans, D.~J., \& Burggraf, B. 2014, \apj, 790, 48 (Paper II)   

\bibitem[Humphreys et al.(2015)]{RMH15}Humphreys, R. M., Martin, J. C., \& Gordon, M. S. 2015, \pasp, 127, 347   

\bibitem[Humphreys et al.(2016)]{RMH16}Humphreys, R. M.,Weis, K., Davidson, K., \&  Gordon, M. S. 2016, \apj, 825, 64  

\bibitem[Jennings et al.(2014)]{Jennings} Jennings, Z. G., Williams, B. F., Murphy, J. W.,  et al. 2014, ApJ., 795, 170 

\bibitem[Kaluzny et al.(1998)]{Kaluzny}Kaluzny, J., Stanek, K. Z., Krockenberger, M. et al. 1998, \aj, 115, 1016 

\bibitem[Kraus et al.(2014)]{Kraus}Kraus, M., Cidale, L. S., Arias, M. L., Oksala, M. E. \& Borges Fernandes, M. 2014, \apjl, 780, L10

\bibitem[Lee, et al.(2014)]{Lee}Lee, C.-H., Seitz, S., Kodric, M., et al. 2014, \apj, 785, 11  

\bibitem[Massey et al.(1995)]{Massey95}Massey, P., Armandroff, T. E., Pyke, R., Patel, K., \& Wilson, C. D. 1995, \aj, 110, 2715
\bibitem[Massey \& Johnson(1998)]{Massey98}Massey, P., \& Johnson, O. 1998, \apj, 505, 793

\bibitem[Massey et al.(2006)]{Massey06}Massey, P., Olsen, K. A. G., Hodge, P. W. et al. 2006, \aj, 131, 2478

\bibitem[Massey et al.(2007)]{Massey07}Massey, P., McNeill, R. T., Olsen, K. A. G., et al. 2007, \aj, 134, 2474 

\bibitem[Massey et al.(2016)]{Massey16}Massey, P., Neugent, K. F., \& Smart, B. M. 2016, \aj, 152, 62  

\bibitem[McQuinn et al.(2007)]{McQ}McQuinn, K. B. W., Woodward, C. E., Willner, S. P. et al. 2007, \apj, 664, 850 

\bibitem[Neugent \& Massey(2011)]{Neugent}Neugent, K. F., \& Massey, P. 2011, \apj, 733, 123

\bibitem[Monteverde et al.(1996)]{Mont}Monteverde, M. L., Herrero, A., Lennon, D. J. et al. 1996, \aap, 312, 24

\bibitem[Mould et al.(2008)]{Mould}Mould, J., Barmby, P., Gordon, K., et al. 2008, \apj, 687, 230  

\bibitem[Oksala et al.(2013)]{Oksala}Oksala, M. E., Kraus, M., Cidale, L. S., Muratore, M. F., \& Borges Fernandes, 2013, \aap, 558, A17 

\bibitem[Sholukhova et al.(2015)]{Shol}Sholukhova, O., Bizyaev, D., Fabrika, S., et al. 2015, \mnras, 447, 2459  
\bibitem[Smartt et al.(2009)]{Smartt09}Smartt, S.~J., Eldridge, J.~J., Crockett, R.~M., \& {Maund}, J.~R. 2009, \mnras, 395, 1409

\bibitem[Smartt et al. (2015)]{Smartt15}Smartt,S.~J.2015, \pasa, 32, 16 

\bibitem[Szeifert et al.(1996)]{Szeif}Szeifert, T., Humphreys, R. M., Davidson, K., et al. 1995, \aap, 314, 131


\bibitem[Thompson et al.(2009)]{Thom}Thompson, T.A., Prieto, J. L., Stanek, K. Z. et al. 2009, \apj, 705, 1364 

\bibitem[Valeev et al.(2010)]{Valeev}Valeev, A. F., Sholukhova, O. N., \& Fabrika, S. N. 2010, Astrophysical Bull., 65, 140 

\bibitem[Wright et al.(2010)]{Wright}Wright, E. L. Eisenhardt, P.R. M., Mainzer, A. K., et al. 2010, \aj, 140, 1868 
\end{thebibliography}
\end{document}